\documentclass[twocolumn,floatfix, nofootinbib,prd,preprintnumbers,showpacs,showkeys,superscriptaddress,longbibliography]{revtex4-1}

\usepackage{coffeestains}
\usepackage[mathscr]{euscript}
\usepackage{amsmath}
\usepackage{graphicx}
\usepackage{dcolumn}
\usepackage{bm}
\usepackage{epsfig}
\usepackage{amssymb,latexsym,mathrsfs}
\usepackage{graphicx}
\usepackage{color}
\usepackage{hyperref}
\usepackage{float}
\usepackage{diagbox}

\usepackage{tikz-cd}
\usetikzlibrary{arrows,decorations.pathmorphing,backgrounds,positioning,fit,matrix}
\usepackage{amsmath}
\usepackage{physics}
\usepackage{csquotes} 
\newcommand*\diff{\mathop{}\!\mathrm{d}}

\hypersetup{
    colorlinks=true,
    linkcolor=red,
    citecolor=blue,
}

\usepackage{orcidlink}

\usepackage{amsmath}
\usepackage{amssymb}
\usepackage{subfigure}
\usepackage{hyperref}
\usepackage{url}
\usepackage{xcolor}
\usepackage{color}
\definecolor{amaranth}{rgb}{0.9, 0.17, 0.31}
\definecolor{purple(munsell)}{rgb}{0.62, 0.0, 0.77}
\definecolor{americanrose}{rgb}{1.0, 0.01, 0.24}
\definecolor{palatinateblue}{rgb}{0.15, 0.23, 0.89}
\definecolor{royalblue(web)}{rgb}{0.25, 0.41, 0.88}
\definecolor{hanpurple}{rgb}{0.32, 0.09, 0.98}
\definecolor{beaublue}{rgb}{0.74, 0.83, 0.9}
\definecolor{carminered}{rgb}{1.0, 0.0, 0.22}
\definecolor{brightpink}{rgb}{1.0, 0.0, 0.5}
\definecolor{vividviolet}{rgb}{0.62, 0.0, 1.0}
\hypersetup{ linktoc=all,
    colorlinks, linkcolor={palatinateblue},
    citecolor={brightpink}, urlcolor={amaranth}}

\newcommand{\be}{\begin{equation}}
\newcommand{\ee}{\end{equation}}
\newcommand{\bs}{\begin{split}} 
\newcommand{\bea}{\begin{eqnarray}}
\newcommand{\eea}{\end{eqnarray}}

\newcommand{\ievlev}[1]{\textcolor{blue}{[{\bf EI}: #1]}}

\newcommand{\bes}{\begin{subequations}}
\newcommand{\ees}{\end{subequations}}

\newcommand{\bo}{\raise-1mm\hbox{\Large$\Box$}}

\newcommand{\bd}{\boldsymbol}

\begin{document}

\preprint{FTPI-MINN-24-11}

\title{Electron-mirror duality and thermality}
\author{Evgenii Ievlev}
\email{ievle001@umn.edu}
\affiliation{William I. Fine Theoretical Physics Institute, School of Physics and Astronomy,\\
University of Minnesota, Minneapolis, MN 55455, USA}
\author{Michael R.R. Good}
\email{michael.good@nu.edu.kz}
\affiliation{Department of Physics \& Energetic Cosmos Laboratory,\\Nazarbayev University,
Astana 010000, Qazaqstan}
\affiliation{Leung Center for Cosmology and Particle Astrophysics,\\
National Taiwan University, Taipei 10617, Taiwan}
\author{Paul C.W. Davies}
\email{paul.davies@asu.edu}
\affiliation{Department of Physics and Beyond: Center for Fundamental Concepts in Science,\\
Arizona State University, Tempe, Arizona 85287, USA}

\begin{abstract} 
Classical electromagnetic radiation from moving point charges is foundational, but the thermal dynamics responsible for classical acceleration temperature are poorly understood.  We investigate the thermal properties of classical electromagnetic radiation in the context of the correspondence between accelerated electrons and moving mirrors, focusing on three trajectories with asymptotically infinite (Davies-Fulling), asymptotically zero (Walker-Davies), and eternally uniform acceleration. The latter two are argued not to be thermal, while the former is found to emit thermal photons with a temperature that depends on the electron's speed. Thermal radiation from the mirror reveals a zero-jerk condition. 

%
%

\end{abstract} 

\keywords{moving mirrors, black hole evaporation, acceleration radiation, Larmor power}
\pacs{41.60.-m (Radiation by moving charges), 05.70.-a (Thermodynamics),
04.70.Dy (Quantum aspects of black holes),
04.62.+v (Quantum field theory in curved spacetime)}
\date{\today} 

\maketitle

\tableofcontents


\section{Introduction}

One of the most discussed results in quantum field theory is the thermal nature of acceleration \cite{Davies:1974th,unruh76}, rendered all the more significant given its close analogy with black hole radiance \cite{Hawking:1974rv,Hawking:1974sw}. It has long been speculated that ‘acceleration thermality’ is a pointer to some profound link between gravitation, spacetime symmetries, holography, and the quantum vacuum so that accelerating systems might provide a window into quantum gravity. For example, Penrose has used this link to argue for gravitationally-induced wave function collapse \cite{Penrose:1996cv}. However, despite decades of speculation, the nature of such a link remains obscure.

To revisit the possibility of a deeper connection, we point out in this paper that the thermal properties of acceleration extend beyond the traditional Unruh and DeWitt particle detector models \cite{unruh76, dewitt_detector} to include accelerating mirrors \cite{moore1970quantum,Fulling:1972md,Davies:1976hi,Davies:1977yv} and classical point charges \cite{Unruh:1982ic,Ford:1982ct,Nikishov:1995qs}. We also present some remarkable concordances in the predicted spectra and energy fluxes among physically very different systems, hinting that a deep principle may be at work.

Because of the formidable difficulty of directly observing Hawking radiation, there has been a growing interest in black hole analogs for ‘table-top’ laboratory testing . 
We also note that several recent experiments probe accelerating mirror effects, see e.g., \cite{AnaBHEL:2022sri, Chen:2015bcg,Chen:2020sir, Steinhauer:2014dra}. 
We feel that broadening the context of the thermal connection with acceleration opens up new possibilities for such analog laboratory tests, see e.g., \cite{Lynch:2022rqx,RDKII:2016lpd,nico}.

Indeed, Larmor's classical electromagnetic acceleration radiation from moving point charges \cite{Larmor1897} is intimately connected to the quantum scalar radiation of the moving mirror \cite{DeWitt:1975ys,Davies:1976hi,Davies:1977yv}.
The first hints of this relationship between electrons and mirrors were pioneered in 1982 by Unruh-Wald \cite{Unruh:1982ic} and Ford-Vilenkin \cite{Ford:1982ct}. In 1995  Nikishov-Ritus \cite{Nikishov:1995qs} matured the particle correspondence significantly, laying the groundwork for subsequent research, see e.g., \cite{Ritus:1999eu,Ritus:2002rq,Ritus:2003wu,Zhakenuly:2021pfm,Ritus:2022bph,Ievlev:2023inj,Ievlev:2023bzk,Good:2022eub,Ievlev:2023inj,Good:2022xin,Lynch:2022rqx}.

The 1+1d moving mirror model particle creation is characterized by the beta Bogolubov coefficients $\beta^R_{pq}$ with $p$, $q$ being the in- and out-scalar mode frequencies\footnote{We use the notation  $p$, $q$ to avoid confusing the scalar frequencies with the 3+1d photon frequency $\omega$. The subscript $R$ refers to the half-space to the right of the mirror on the line.}, for an introduction see e.g. \cite{good2013time} and the textbooks \cite{Birrell:1982ix,Parker:2009uva}.
On the other hand, a classical charge $e$ moving in a rectilinear fashion produces radiation with spectral distribution $\frac{\diff{I}}{\diff{\Omega}}(\omega,\cos\theta)$ depending on the frequency $\omega$ and the observation angle $\theta$, $\Omega$ being the solid angle. 

The recipe for converting between the electron and the mirror is given by a simple relationship between the spectral distribution on the electron side and the Bogolubov coefficient squared on the mirror side \cite{Ievlev:2023bzk,Ievlev:2023inj}:
\begin{equation}
\begin{aligned}
	&\frac{\diff{I}}{\diff{\Omega}}(\omega,\cos\theta) = \frac{e^2 \omega^2}{4\pi} |\beta^R_{pq}|^2, \\
	&p + q = \omega \,, \quad p - q = \omega \cos\theta, \\
	&p = \omega \frac{1 + \cos\theta }{2} \,, \quad 	q = \omega \frac{1 - \cos\theta }{2}.
\end{aligned}
\label{recipe_dIdOmega_from_mirror}
\end{equation}
Here, the electron moves in three-dimensional space along a straight line (say, $z$-axis) with the same time dependence $z(t)$ as the mirror on the line.
This mapping allows accelerated electrons to be exploited as bona fide black hole radiation analogs similar to moving mirrors, whose history as evaporation analogs \cite{Holzhey:1994we,wilczek1993quantum,walker1985particle,Walker:1984ya,CW2lifetime} have been fruitful; see also \cite{ Ievlev:2023ejs,Osawa:2024fqb,Lin:2021bpe,Kumar:2023kse,Reyes:2021npy} for some recent works.

This mapping contrasts with the fact that acceleration-thermality has been primarily studied in the context of quantum theories, mainly through the uniform acceleration Davies-Unruh effect \cite{Davies:1974th,unruh76}, which has helped develop connections to the classical radiation of Larmor, see e.g., \cite{Cozzella:2020gci,Landulfo:2019tqj,doi:10.1098/rspa.2020.0656,Cozzella:2017ckb}.
A classical connection between acceleration and temperature (classical acceleration temperature, or `CAT'), as observed in the Planck electromagnetic radiation spectrum, has largely remained unexplored until recent efforts \cite{Good:2022eub,Ievlev:2023inj,Good:2022xin,Ievlev:2023bzk,Ievlev:2023akh}.
Classical systems are often more intuitive than quantum ones, and tractability allows the analysis of different radiative thermodynamics due to non-uniform acceleration.
For instance, it was recently established that non-uniformly accelerated charges can produce electromagnetic radiation with a thermal spectrum \cite{Good:2022eub}; see a recent theory paper \cite{Ievlev:2023inj} and experimental evidence from beta decay \cite{Lynch:2022rqx}.

In this work, we present new results on the thermal properties of classical electromagnetic radiation and its connection to the quantum radiation from moving mirrors.
We focus on three notable examples: the Davies-Fulling trajectory (infinite asymptotic acceleration), the Walker-Davies trajectory (zero asymptotic acceleration), and the eternal-uniform trajectory (constant asymptotic acceleration).
We investigate the radiation from a classical point charge moving along these trajectories and find that while the first example exhibits thermality, the second and third do not.
On the mirror side, this turns out to be connected to the zero jerk condition (jerk being the time derivative of the acceleration), as explained below.

The structure of the paper is as follows.
In Sec.~\ref{sec:DF}, we review the Davies-Fulling trajectory and its essential dynamics, developing an intuition for the power and the energy emission. We compute the scalar/photon spectra, and specialize to slow speeds. Sec.~\ref{sec:WD} is devoted to the Walker-Davies trajectory, where we perform spectral analysis of the radiation and systematic computation in analog to Sec.~\ref{sec:DF}.  
In Sec.~\ref{sec:UA}, we compare the previous spectral results with the proper uniformly accelerated trajectory. 
Sec.~\ref{sec:CONCLUSIONS} summarizes the main findings, emphasizing the speed-dependence of the Davies-Fulling trajectory's temperature and the Walker-Davies trajectory particle count. 
Appendix \ref{Appendix:UA} and \ref{Appendix:JERK} present details on the spectral distribution of uniform acceleration and zero-jerk thermality, respectively. We use units $k_\textrm{B} = c = \mu_0 = \epsilon_0 =1$, where the fine structure constant in terms of electric charge is $\alpha_{\textrm{fs}} = e^2/4\pi \hbar \approx 1/137$. For classical/quantum clarity, temperatures derived classically are expressed in the Stoney scale (no $\hbar$ \cite{stoney1881physical}; see also \cite{Ievlev:2023inj, barrowSTONE}).  Quantum temperatures are expressed in the Kelvin scale ($\hbar$ required \cite{SI}).

\section{Davies-Fulling trajectory}
\label{sec:DF}


The 1977 Davies-Fulling trajectory \cite{Davies:1977yv}, $\dot{z} = -\tanh t$, illuminated the thermodynamic duality between black holes and moving mirrors.  The late-time thermal behavior describes the nature of equilibrium black hole evaporation \cite{Hawking:1974sw}.  Here, we wish to understand the radiation emitted from an electron that travels along this particular globally-defined worldline.

\subsection{Dynamics and total energy}
\begin{figure}[htbp]
\centering
  \centering
  \includegraphics[width=0.9\linewidth]{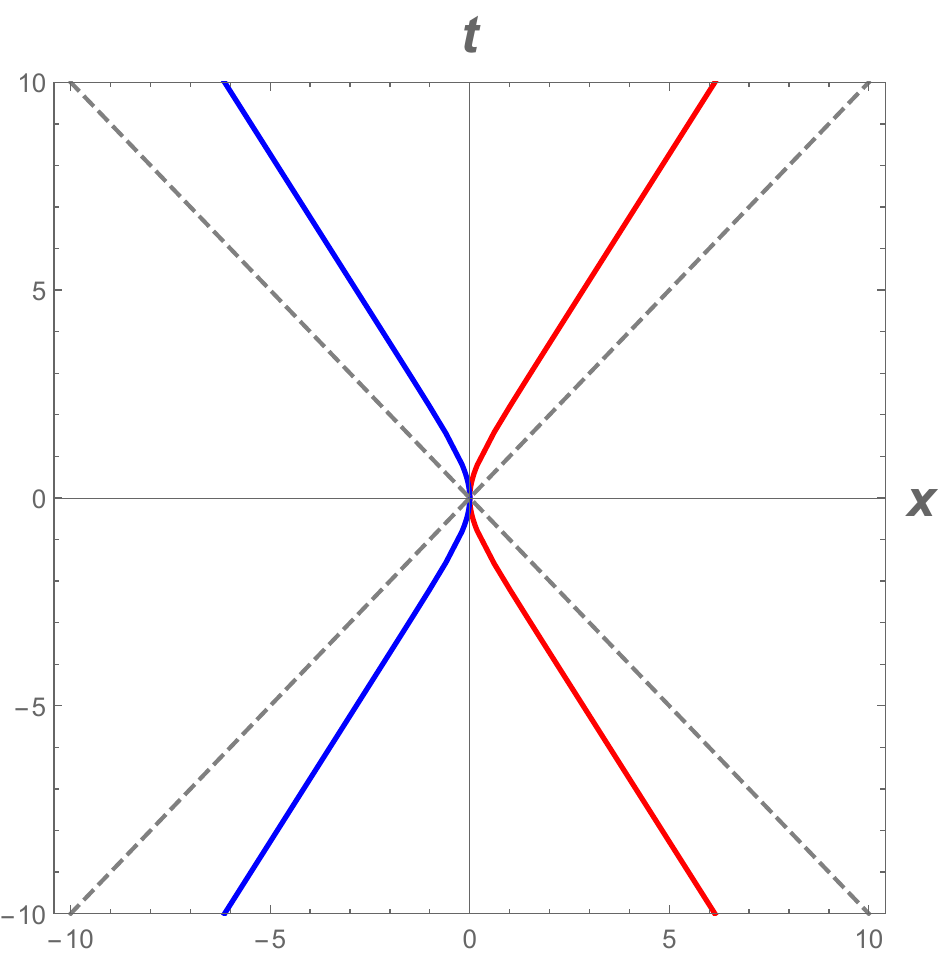}
 \caption{A spacetime diagram of Eq.~(\ref{eom}) (red), demonstrating qualitative hyperbolic shape and time-reversal symmetry. Here, the final speed is set to $s= 0.66$ and $\kappa = 1$.  The red trajectory is a parity flipped Eq.~(\ref{eom}), added for illustrative symmetry.  Interestingly, the peel acceleration $\kappa$ goes constant when $s \to 1$ in the upper-left and lower-right quadrants.  Thermal radiation is more likely when the trajectory is asymptotically moving to the left, away from the $\mathscr{I}^+_R$ observer. }
\label{Spacetime}
\end{figure}

\begin{figure}[htbp]
\centering
  \centering
  \includegraphics[width=0.9\linewidth]{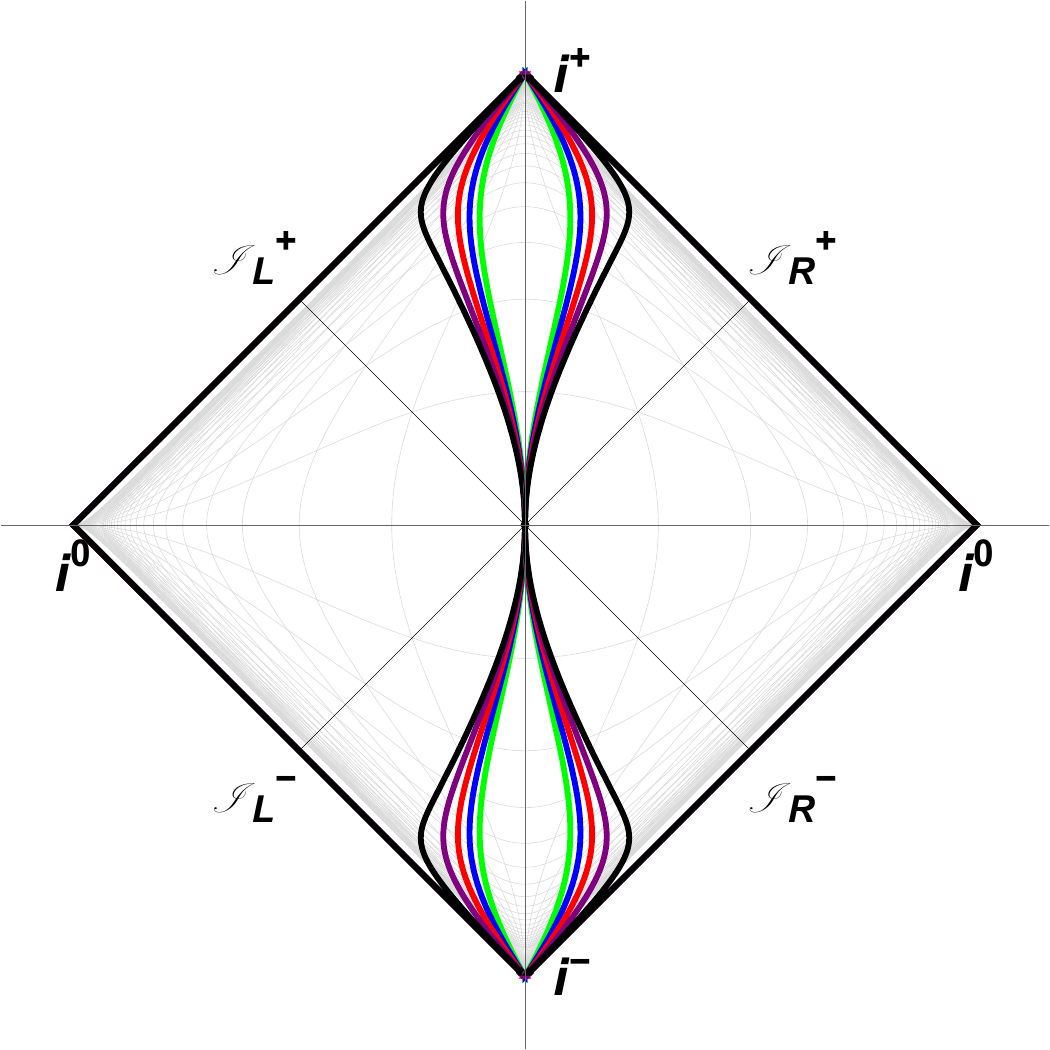}
 \caption{A Penrose conformal diagram of Eq.~(\ref{eom}), demonstrating asymptotic inertia of $\alpha(t)\to 0$ as $t\to \pm\infty$ because final speed is $s<1$ resulting in initial and final states approaching timelike future infinity. From the inside-out, the final speeds are $s=0.55,0.65,0.75,0.85,0.95$. Here, $\kappa = 1$, and trajectories are plotted traveling both left and right for illustrative symmetry. }
\label{Penrose}
\end{figure}

Consider the electron moving along the Davies-Fulling trajectory \cite{Davies:1977yv} 
\be z(t) = -\frac{s}{\kappa}\ln \cosh \kappa t,\label{eom}\ee
with an asymptotic approach to constant speed $s$ such that the system is asymptotically inertial \cite{Good:2017kjr}. See Figure \ref{Spacetime} for a spacetime plot (blue line) and Figure \ref{Penrose} for a conformal diagram of the trajectory class.  Here velocity, $v(t)=\dot{z}(t)$, and rapidity, $\eta= \tanh^{-1}v$, are given by,
\be
v(t)= - s \tanh (\kappa  t),
\qquad 
\eta(t)= - \tanh^{-1}[s \tanh (\kappa  t)].
\ee
Notice that at ultra-relativistic final speeds, the rapidity becomes linear in time, $\eta(t) \sim - \kappa t$.

Seeing as the Lorentz factor is given by $\gamma = \cosh \eta$, 
one can compute the rectilinear proper acceleration, $\alpha(t)=\gamma^3\ddot{z}(t)$, straightforwardly
\be
\alpha(t)= - \frac{s~\kappa~\text{sech}^2(\kappa  t)}{\left(1-s^2 \tanh ^2(\kappa  t)\right)^{3/2}}.
\ee
But what we are really after is the `peel' acceleration \cite{CW2lifetime}, $\mathcal{P}(t) = 2\alpha e^\eta$, which is thermodynamically relevant \cite{Ievlev:2023bzk,Ievlev:2023inj,Good:2022eub,Bianchi:2014qua,Barcelo:2010pj}.  In the case of the worldline, Eq.~(\ref{eom}), the peel is given by,
\begin{equation}
    \mathcal{P}(t) = - \frac{2 \kappa  s~ \text{sech}^2(\kappa  t)}{(1 + s \tanh (\kappa  t))^2 (1 - s \tanh (\kappa  t))} \,.
\label{DF_peel}
\end{equation}
In the regime where $s \to 1$ the peel simplifies to $\mathcal{P}(t) = - (1 + e^{- 2 \kappa t}) \kappa$.
From this, we see that at late times, $t \to \infty$, the peel approaches a constant, $\mathcal{P}(t) \to - \kappa$. 
This suggests thermal emission will occur in the far future. See the blue worldline electron trajectory of Figure \ref{Spacetime} moving to the left on the upper-left-hand side of the spacetime diagram. This is where we expect thermal emission.

\paragraph*{Force, power, energy}
A moving point charge experiences a radiation reaction. The magnitude of the self-force is
\be F = \frac{e^2\alpha'(\tau)}{6\pi}, \quad \lim_{s\to 1}F = -\frac{e^2\kappa^2}{12\pi}\sinh{2\kappa t},\ee
where $\tau$ is proper time, and the prime indicates a derivative with respect to the argument.  The second expression is valid for final ultra-relativistic speeds.
The total Larmor power radiated by the point charge as it accelerates is given by,
\be P= \frac{e^2\alpha^2}{6\pi} = \frac{e^2 s^2~\kappa^2~\text{sech}^4(\kappa  t)}{6 \pi\left(1-s^2 \tanh ^2(\kappa  t)\right)^{3}}. \label{Power}\ee
It is possible to find the total energy via the self-force or the power:
\be E_{\textrm{electron}} = \int_{-\infty}^{\infty} P \diff{t} = -\int_{-\infty}^{\infty} F\cdot v \diff{t}.\ee
The result for the total energy emitted by an electron along the Davies-Fulling trajectory, Eq.~(\ref{eom}), is analytic:
\be E_{\textrm{electron}} = \frac{e^2\kappa}{24\pi}\left[2\gamma_s^2-3 + \left(4-\frac{3}{\gamma_s^2}\right)\frac{\eta_s}{s}\right].\label{totalenergy}\ee
Here, the final Lorentz factor is $\gamma_s = \cosh \eta_s$.  The final speed $s$ is given by the final rapidity $s = \tanh \eta_s$. This analytic form of the energy will help us confirm the consistency of the following spectral analysis, which is needed to prove thermal radiation. 


\subsection{Electron and mirror spectra}

Having discussed the Davies-Fulling trajectory's time-dependent properties, let us now turn to the spectrum of the electron's radiation and its connection to the mirror moving on the same trajectory.

\paragraph{Bogolubov Spectra for the mirror}

Treating the electron as a tiny double-sided moving mirror (e.g., \cite{Good:2022eub,Ievlev:2023inj,Ievlev:2023bzk} and references therein), gives the 
$\beta^R_{pq}$ spectrum \cite{Davies:1977yv,Good:2017kjr},
\begin{equation}
    \beta^R_{pq} = \frac{2^{i s (p-q) / \kappa} \sqrt{pq} }{ 2 \pi \kappa (p-q) } \, 
        B\left( \frac{i g_- }{2\kappa}  , \, \frac{i g_+ }{2\kappa}  \right) \,,
\label{DF_beta_rhs}
\end{equation}
where $g_\pm = s (p-q) \pm (p+q)$.
The double-sided spectrum is defined as  $|\beta_{pq}|^2 =  |\beta_{pq}^R|^2 +  |\beta_{pq}^L|^2$, and with $|\beta_{pq}^L|^2=|\beta_{qp}^R|^2$ we obtain
\be |\beta_{pq}|^2 = \frac{4 s pq \sinh \left[\frac{\pi  s (p-q)}{\kappa }\right]}{\pi  a b \kappa  (p-q) \left(\cosh \left[\frac{\pi (p+q)}{\kappa }\right]-\cosh \left[\frac{\pi  s (p-q)}{\kappa }\right]\right)}.\label{beta2}\ee
Here $a= p(1+s) + q(1-s)$, $b=p(1-s)+q(1+s)$.  
A numerical integration recapitulates the total energy of the emitted radiation:
\be E_{\textrm{mirror}} = \int_0^\infty \int_0^\infty \hbar p |\beta_{pq}|^2 \diff{p}\diff{q}.\label{totE}\ee
That is, one can be confident Eq.~(\ref{beta2}) is consistent with the Larmor form because Eq.~(\ref{totE}) gives the analog version of Eq.~(\ref{totalenergy}), $ E_{\textrm{mirror}}/\hbar = E_{\textrm{electron}}/e^2$.

\paragraph{Thermal limit of the mirror's radiation}
Interestingly, in the limit of the speed of light, $s\to 1$, the beta Bogolubov spectrum, Eq.~(\ref{beta2}), simplifies to (see also \cite{Fernandez-Silvestre:2022gqn})
\be |\beta_{pq}|^2 =\frac{1}{\pi \kappa (p-q)} \left[\frac{1}{e^{2 \pi  q/\kappa}-1}-\frac{1}{e^{2 \pi  p/\kappa}-1}\right].\label{DFb2}
\ee
which is the light-speed case \cite{Good:2017ddq} (see also \cite{Fernandez-Silvestre:2022gqn}) of the drifting counterpart in \cite{Good:2017kjr}, first suggested by Davies-Fulling \cite{Davies:1977yv} and studied at late-times, see Birrell-Davies \cite{Birrell:1982ix}. 
Consider that Eq.~(\ref{DFb2}) in the high frequency approximation, $q \gg p$, gives
\be |\beta_{pq}|^2 =\frac{1}{\pi \kappa q}\frac{1}{e^{2 \pi  p/\kappa}-1}.\label{highDFb2}
\ee
while the low frequency approximation, $q \ll p$, gives
\be |\beta_{pq}|^2 =\frac{1}{\pi \kappa p} \frac{1}{e^{2 \pi  q/\kappa}-1}.\label{lowDFb2}
\ee
These explicit Planck distributions accurately characterize the moving mirror radiation (scalars) with a temperature, $T = \kappa/2\pi$.  They also suggest exploring the regime of high speeds $s \to 1$ for study with respect to photon radiation from the corresponding electron.

\paragraph{Spectral distribution for the electron}

The trajectory is in Eq.~\eqref{eom}.
To calculate the spectral distribution of the electron's radiation, we will use the formula (see Eq. (14.70) of \cite{Jackson:490457}):
\begin{equation}
    \frac{\diff I(\omega)}{\diff \Omega} 
		= \frac{\omega^2}{16 \pi^3} \sin^2\theta \, \abs{   j_z(\omega, k_z ) }^2 
\label{I_Omega_Jackson}
\end{equation}
Here, $j_z(\omega, k_z )$ is the Fourier transform of the current,
\begin{equation}
\begin{aligned}
	j_z (\omega, k_z) 
		&= e\, \int\limits_{- \infty}^{\infty} \diff{t} \, \dv{z}{t} \, e^{-i (\omega t - k_z z(t)) } \\
		&=  \frac{ e \omega}{k_z}\,  \int\limits_{- \infty}^{\infty} \diff{t}  \, e^{-i (\omega t - k_z z(t)) } \,.
\end{aligned}
\label{jz_fourier_transform_definition}
\end{equation}
Here $k_z = |\bd{k}|\cos\theta = \omega \cos\theta$. The second line can be obtained by integrating parts or using the charge conservation law in Fourier space.
%
Substituting here the trajectory from Eq.~\eqref{eom} and making a change of variables $e^{2\kappa t} = w$, $w = x / (1-x)$, the integral can be brought to the Euler beta function form with the result
\begin{equation}
    j_z (\omega, k_z) = \frac{e \omega 2^{i \frac{s k_z}{\kappa}}}{2\kappa k_z } \, 
			B\left( - i \frac{\omega - s k_z}{2 \kappa} \,, \ i \frac{\omega + s k_z}{2 \kappa} \right)
\label{jz_fourier_transform_ans_1}
\end{equation}
Plugging this into Eq.~\eqref{I_Omega_Jackson} and computing the absolute value through gamma functions, we arrive at the spectral distribution
\begin{equation}
\begin{aligned}
	\frac{\diff I(\omega)}{\diff \Omega}
		&= \frac{ e^2 s  \sin^2\theta }{ 8  \pi^2  \cos\theta (1 - s^2 \cos^2\theta ) } \\
				&\times \frac{\omega}{\kappa}
				\left[ 
					\frac{1}{e^{ \pi \omega \frac{1 - s \cos\theta }{\kappa} } - 1}
					- \frac{1}{e^{ \pi \omega \frac{1 + s \cos\theta }{\kappa} } - 1}
				\right]
\end{aligned}
\label{dIdOmega_3}
\end{equation}
Plugging the results for the mirror Eq.~\eqref{DF_beta_rhs} and the electron Eq.~\eqref{dIdOmega_3} in the two sides of the electron-mirror correspondence Eq.~\eqref{recipe_dIdOmega_from_mirror} we have checked that this correspondence indeed works. Computing the total energy emitted by the electron using Eq.~\eqref{dIdOmega_3} gives Eq.~\eqref{totalenergy}.

\begin{table}[ht] 
\begin{tabular}{l | l }
Mirror & \ Electron   \\ \hline
\rule{0pt}{1.05\normalbaselineskip}moves to the left\ \ & \ moves down $z \to - \infty$  \\
	high-freq.\ $q \gg p$ & \ blueshift-forward $\theta \to \pi$, $q \approx \omega$ \\
	low-freq. $q \ll p$ & \ redshift-recede $\theta \to 0$, $p \approx \omega$
\end{tabular} \\ 
\caption{Correspondence between mirror and electron properties in various limits.} 
\label{tab:limits} 
\end{table}

\paragraph{Thermal limits}

Let's look at the redshift-receding limit $\theta \sim 0$, see Table~\ref{tab:limits} for comparison of the corresponding limits in the moving mirror and the point charge problems.
At high speeds $s \to 1$, the first term in the square brackets in Eq.~\eqref{dIdOmega_3} dominates. We have\footnote{At very low frequencies $\omega \lesssim \omega_\text{IR} = \kappa / (1+s)$ the approximation Eq.~\eqref{DF_dIdOmega_redshift} does not work; however, in the relativistic limit the corresponding characteristic frequency $\omega_\text{IR}$ is negligible compared to the temperature Eq.~\eqref{T_red}. The same goes for the blueshift-forward limit discussed below. }
\begin{equation}
    \frac{1}{ \sin^2\theta } \frac{\diff I(\omega)}{\diff \Omega} \approx \frac{ e^2 s    }{ 8  \pi^2 (1 - s^2  ) } \, 
						\frac{\omega}{\kappa}
						\frac{  1 }{ e^{ \pi \omega (1-s) / \kappa } - 1 }
\label{DF_dIdOmega_redshift}
\end{equation}
%
%
This has the form of a 1+1D Planck distribution.
The corresponding temperature is
\begin{equation}
	T = \frac{e^2\kappa}{\pi (1-s)} \qquad \textrm{(Classical; Stoney Scale)}
\label{T_red}
\end{equation}
Note that $s > 0$ in this formula.
%

In the opposite (\textquote{blueshift-forward}) limit $\theta \sim \pi$ it is the second term in the square brackets in Eq.~\eqref{dIdOmega_3} that contributes, and the distribution is again Planck with the same temperature as in Eq.~\eqref{T_red}.

\paragraph{Discussion: why is it so hot?}

One may wonder why the temperature in Eq.~\eqref{T_red} is so hot compared to the results for the mirror. 
On the mirror side, in the low-frequency regime $q \ll p$, the particle distribution reduces to a Planck factor Eq.~\eqref{lowDFb2}.
%
From this, one can conclude that the temperature of the radiation from this mirror is 
\begin{equation}
	T_\text{mirror} = \frac{\hbar\kappa}{2\pi}  \qquad \textrm{(Quantum; Kelvin Scale)} \label{mirrortemp}
\end{equation}
However, this is a temperature for the frequency $q$!

The electron-mirror correspondence Eq.~\eqref{recipe_dIdOmega_from_mirror} tells us that the frequencies on the two sides are not identical; rather, they are related via
\begin{equation}
	p = \frac{\omega (1 + \cos\theta)}{2} \,, \quad
	q = \frac{\omega (1 - \cos\theta)}{2}
\label{recipe_pq_1}
\end{equation}
The \textquote{low-frequency} regime $q \ll p$ then corresponds to the  $\theta \sim 0$, cf. Table~\ref{tab:limits}.

With the recipe Eq.~\eqref{recipe_pq_1} the distribution for the mirror Eq.~\eqref{lowDFb2} transforms to the distribution for the electron Eq.~\eqref{DF_dIdOmega_redshift} with the correct temperature Eq.~\eqref{T_red}.
The result is the same for the \textquote{high-frequency} mirror $p \gg q$.

It is amazing how the meaning of temperature might change when we pass from a mirror to an electron.

\subsection{Non-relativistic speeds}
For a given $\kappa$, the electron can be much hotter than the mirror when $s\approx 1$ as seen by Eq.~(\ref{T_red}).  
However, the electron is comparable to the mirror when $s\approx 0$ (also in the redshift receding limit).  For small speeds, the electron temperature, from Eq.~(\ref{dIdOmega_3}) or Eq.~(\ref{T_red}) and mirror temperature, Eq.~(\ref{mirrortemp}), are respectively,
\be T_{\textrm{mirror}} = \frac{T_{\textrm{electron}}}{2},\quad (\textrm{Kelvin Scale}).\ee
The energy Eq.~(\ref{totalenergy}) also simplifies to leading order,
\be E_{\textrm{electron}} = \frac{e^2}{\hbar} E_{\textrm{mirror}}  = \frac{2 e^2 \kappa }{9\pi} s^2,\ee
in the non-relativistic limit, $s\approx 0$. 

\section{Walker-Davies trajectory}\label{sec:WD}
The first asymptotically resting trajectory with analytically solved Bogolubov coefficients was found in \cite{Walker_1982}:
\begin{equation}
	t(z) = - z \pm A \sqrt{ e^{- 2 z / B} - 1 }.
\label{wd_traj}
\end{equation}
The parameters $A > B > 0$ are related to the maximum speed of the motion. Eq.~\eqref{wd_traj} with the $-$ sign parametrizes the trajectory in the interval $t \in (-\infty, 0]$, while Eq.~\eqref{wd_traj} with the $+$ sign gives the trajectory for $t \in [0, + \infty)$. Lightcone coordinates with retarded time $U = t-z$ and advanced time $V = t+z$, conveniently expresses the Walker-Davies trajectory, Eq.~(\ref{wd_traj}) as
\be U(V) = V+ B \ln \left(V^2/A^2+1\right). \label{wd_traj_lightcone} \ee
In the following, we will solve for the emission properties of an electron that travels along this worldline.

\subsection{Dynamics and total energy}

In terms of the Larmor power, the total energy emitted by the electron along its worldline is
\begin{equation}
\begin{aligned}
	E_\text{tot}^{e} &= \int\limits_{-\infty}^{\infty} \diff t \, P(t) \,, \quad
	P(t) = \frac{e^2\alpha^2(t)}{6 \pi} \,, \\
	\alpha(t) &= \gamma^3 \ddot{z}(t) \,, \quad
	\gamma = \frac{1}{\sqrt{1-v^2}}
\end{aligned}
\label{Etot_e-Larm}
\end{equation}
This formula can be adapted for use with a trajectory in the form $t = t(z)$:
\begin{equation}
\begin{aligned}
	v(z) &= \dot{z} = \left( \dv{t}{z} \right)^{-1}, \\
	E_\text{tot}^{e} &= \int\limits_{-\infty}^{\infty} \diff z \,  
		\frac{ e^2 |v(z)| }{6 \pi (1 - v^2(z))^3 }
		\left( \dv{v(z)}{z} \right)^2,
\end{aligned}
\end{equation}
%
%
Substituting the Walker-Davies trajectory Eq.~\eqref{wd_traj} and carefully evaluating the above integral (which is not particularly straightforward), we arrive at the result
\begin{equation}
	\frac{E_\text{tot}^{e}}{e^2} 
		=  \frac{ B^2 }{ 48 (A^2 - B^2)^{3/2} } 
		+  \frac{1}{ 24 \sqrt{A^2 - B^2} }
		- \frac{1}{ 24 A }.
\label{wd_total_energy}
\end{equation}
%
According to the electron-mirror correspondence, the total radiated energy for an electron and a mirror are related as
$E_\text{tot}^{e} = e^2 / \hbar \cdot E_\text{tot}^\text{mirror} $, where $e$ is the electron's charge.

The first term of Eq.~\eqref{wd_total_energy} exactly reproduces the energy on the right side of the mirror computed in \cite{Walker_1982}.
Therefore, the second two terms give the energy on the left side of the mirror.

We can parametrize $(A,B)$ in terms of the peel scale $\kappa$ and maximum speed of the electron $v_{\textrm{max}}$ along the Walker-Davies trajectory. 
Taking
\begin{equation}
    A = \frac{\pi}{\kappa} \,, \quad
    B = \frac{2\pi v_{\textrm{max}}}{\kappa(1+v_{\textrm{max}})} \,,
\label{AB_to_kappa_v}
\end{equation}
%
the total energy emitted at non-relativistic speeds and relativistic speeds, respectfully, is
\be E \approx \frac{\kappa e^2}{6\pi}v_{\textrm{max}}^2, \qquad E \approx \frac{\kappa e^2 }{12 \sqrt{2} \pi }\gamma_{\textrm{max}} ^3.\ee

\subsection{Electron and mirror spectra}

The beta Bogolubov coefficients for the trajectory Eq.~\eqref{wd_traj} were computed in \cite{Walker_1982}:
\begin{equation}
\begin{aligned}
	\left|\beta_{pq}^R\right|^2 
                &= \frac{2 A B}{\pi^2} \frac{q}{p + q}
				\sinh(\pi p B) \\
				&\times \left|K_{-1 / 2+i p B}\left[ A ( p + q ) \right]\right|^2 \,.
\end{aligned}
\end{equation}
Here, $K$ is the modified Bessel function of the second kind.

We want to consider an electron in a rectilinear motion along the same trajectory Eq.~\eqref{wd_traj}.
Applying the recipe Eq.~\eqref{recipe_dIdOmega_from_mirror} we obtain the electron's radiation spectral distribution:
\begin{equation}
\begin{aligned}
	\frac{\diff{I}}{\diff{\Omega}}
		&= \frac{e^2 \omega^2 A B (1 - \cos\theta ) }{4\pi^3}
			\sinh(\pi \omega (1 + \cos\theta) B / 2) \\
	&\times \left|K_{-1 / 2+i \omega (1 + \cos\theta) B / 2}\left( \omega A \right)\right|^2 \label{WD_SD}
\end{aligned}
\end{equation}
Integration of this spectral distribution over the angles and frequencies gives the total energy.
We have checked numerically that the total energy from the spectral distribution Eq.~(\ref{WD_SD}) coincides with the explicit result Eq.~\eqref{wd_total_energy} from the Larmor power.
%

\subsection{Non-relativistic speeds}

Consider the limit $B \ll A$. 
As is evident from Eq.~\eqref{AB_to_kappa_v}, this corresponds to the limit when the maximal speed $v_\text{max} \ll 1$.
In this limit, the Bessel function is well-approximated by
\begin{equation}
\begin{aligned}
	K_{-1 / 2+i p B} [ A ( p + q ) ]
		&\approx K_{-1 / 2} [ A ( p + q ) ] \\
		&= \sqrt{ \frac{\pi}{ 2 A ( p + q ) } } e^{ - A ( p + q ) }
\end{aligned}
\end{equation}
In this approximation, the mirror's betas and the electron's spectral distribution have simple forms:
\begin{equation}
	\left|\beta_{pq}^R\right|^2
		\approx \frac{ B}{\pi}  \frac{q}{ (p + q)^2 }  \sinh(\pi p B) e^{ - 2 A ( p + q ) },
\end{equation}
%
%
\begin{equation}
	\frac{\diff{I}}{\diff{\Omega}}
		\approx \frac{e^2 \omega  B (1 - \cos\theta ) }{8\pi^2} \sinh(\pi \omega (1 + \cos\theta) B / 2) e^{ - 2 A \omega }.
\end{equation}
%
%
This allows for the analytical computation of various characteristic quantities.
For the mirror, one finds the particle distribution and the total particle count (to the leading order in $B/A$):
\begin{equation}
\begin{aligned}
	N_p^R &= \int\limits_0^\infty \diff q \, |\beta_{pq}^R|^2 
		\approx B^2 p \left[ (1 + 2 A p) \Gamma(0,2Ap) - e^{- 2 A p} \right], \\
	N_\text{tot}^R &= \int\limits_0^\infty \diff p \, N_p^R 
		\approx \frac{ B^2 }{ 24 A^2 } = \frac{v_{\textrm{max}}^2}{6}.
\end{aligned}\label{vmaxPARTICLE}
\end{equation}
Here, $\Gamma(x,a)$ is the incomplete gamma function.
For the electron, one can compute the spectrum and the angular energy distribution (to the leading order in $B/A$):
\begin{equation}
\begin{aligned}
	I(\omega) &= \int \diff\Omega \, \frac{\diff{I}}{\diff{\Omega}} 
		\approx \frac{e^2 B^2}{6} \omega^2 e^{-2 A \omega}, \\
	E(\Omega) &= \int\limits_0^\infty \diff\omega \, \frac{\diff{I}}{\diff{\Omega}} 
		\approx \frac{ e^2 B^2 }{ 64 \pi A^3 } \sin^2\theta.
\end{aligned}
\end{equation}
Notice that $I(\omega)$ has the form of the Wien's law in 2+1D.
We can ascribe this system the temperature (in the UV sense)
\begin{equation}
    T_{\textrm{WD}} = \frac{e^2}{2 A} = \frac{e^2 \kappa}{2\pi}, \qquad \textrm{(Classical; Stoney Scale)}\,.
\end{equation}
However, this assignment is only valid in the non-relativistic limit and the sense of Wien's law rather than Planck's.

The total radiated energy from the electron is particularly simple in this approximation:
\begin{equation}
	E = \int\limits_0^\infty \diff\omega \, I(\omega) 
            = \int \diff\Omega \, E(\Omega)
		\approx \frac{e^2 B^2}{ 24 A^3 },
\end{equation}
and agrees with Eq.~\eqref{wd_total_energy} to leading order in $B\ll A$.

\section{Uniform proper-acceleration trajectory}
\label{sec:UA}

An uncountable number of papers have been written on a uniformly accelerated charge.
There is some controversy concerning whether or not the radiation from such a charge is thermal in some sense.
In this section, we determine the trajectory's spectral distribution, and in the next section, we argue that the radiation's temperature is undefined.

\subsection{Electron and mirror spectra}
We start with known results for the uniformly accelerated mirror.
The beta Bogolubov coefficients for such a mirror are known (see e.g., \cite{Davies:1977yv, good2020extreme,Birrell:1982ix}):
\begin{equation}
	\left|\beta^R_{pq}\right|^2
		= \frac{1}{\pi^2 \kappa^2}\left|K_1\left(\frac{2}{\kappa} \sqrt{pq}\right)\right|^2
\end{equation}
Using the electron-mirror correspondence Eq.~\eqref{recipe_dIdOmega_from_mirror} we instantly obtain the spectral distribution for an electron moving with uniform proper acceleration:
\begin{equation}
    \frac{\diff{I}}{\diff{\Omega}} = \frac{e^2 \omega^2}{4\pi^3 \kappa^2 } \left|K_1\left(\frac{\omega }{\kappa} \sin\theta \right)\right|^2
\label{recipe_dIdOmega_from_mirror_uniform}
\end{equation}
%
From the known asymptotic behavior of the Bessel function
\begin{equation}
	K_\alpha(z) \approx
		\begin{cases}
			\frac{\Gamma(\alpha)}{2} \left( \frac{2}{z} \right)^\alpha \,, \quad &z \to 0\\
			\sqrt{\frac{\pi}{2 z}} e^{-z} \,, \quad &z \to \infty
		\end{cases}
\label{K_limits}
\end{equation}
we can derive asymptotic expressions of the spectral distribution:
\begin{equation}
	\dv{I}{\Omega}() (\omega, \theta) \approx
		\begin{cases}
			\dfrac{e^2 }{16 \pi^3 \sin^2\theta } \,, \quad &\omega \to 0 \\[10pt]
			\dfrac{e^2 \omega}{8 \pi^2 \kappa \sin\theta }  e^{-  \frac{4 \omega }{\kappa} \sin\theta } \,, \quad &\omega \to \infty
		\end{cases}
\label{I_uni_limits}
\end{equation}
Note that a tremendous amount of energy is radiated forward and backward.

%
%
%

\subsection{Would-be thermality}

Let us cast doubt on the thermality of the uniformly proper accelerated electron.
Consider some spectral distribution of a 1+1 Planck form,
\begin{equation}
	f( \omega ) = C \frac{\omega}{e^\frac{\omega}{T} - 1} \,,
\end{equation}
where $C$ is some constant (in the sense that it does not depend on $\omega$).
The IR and UV limits are
\begin{equation}
	f( \omega ) \approx
		\begin{cases}
			CT \,, \quad &\omega \to 0, \\[10pt]
			C \omega e^{-\frac{\omega}{T}} \,, \quad &\omega \to \infty.
		\end{cases}
\label{Planck_uni_limits}
\end{equation}
Comparing Eq.~\eqref{Planck_uni_limits} with Eq.~\eqref{I_uni_limits}, we see that the spectral distribution for the uniformly accelerated electron \eqref{I_uni_limits} looks thermal in the IR and UV limits separately, but these limits are not consistent.

Moreover, if we look at the UV limit only, we get the temperature
\begin{equation}
	T_{\textrm{UV}} = \frac{e^2\kappa}{4  \sin\theta } \qquad \textrm{(Classical; Stoney Scale)} 
\label{uni_temperature}
\end{equation}
which becomes infinite in the redshift-receding limit.  It turns out that the temperature for the uniformly accelerated electron is undefined.
We discuss this in the following section.

\subsection{What is meant by thermality?}

Let us compare the discussion of thermal properties of the Davies-Fulling trajectory in Sec.~\ref{sec:DF} and the uniformly accelerated case in Sec.~\ref{sec:UA}.
In both cases, the UV tails have a form of Wien's law. Can we say that they are both thermal?



\paragraph{Uniformly accelerated electron is not thermal}
The uniformly accelerated electron has infinite total radiated energy.
This is related to the fact that, because of the singular behavior of the spectral distribution $\diff I / \diff \Omega$ at $\theta \sim 0, \pi$ (cf. Eq.~\eqref{I_uni_limits}), the spectrum $I(\omega)$ diverges. 
Therefore, it makes sense only to consider the spectral distribution $\diff I / \diff \Omega$.

At what angles $\theta$ would the spectral distribution be interesting? Since the electron's trajectory is unbounded, we cannot look at this electron from some arbitrary angle $\theta$. As the electron flies away, at late times, we would see it at an angle $\theta \sim 0$.
We will eventually see the electron moving away from us wherever we are in space.
Therefore, we would like to argue that it only makes sense to consider the spectral distribution $\diff I / \diff \Omega$ at $\theta \sim 0$.

In the above argument, we must be concerned only with the \textit{cumulative} spectral distribution, i.e., we are observing the electron over all of its motion.
It would be different if we only investigated the electron's radiation for a finite time.

Since we established the need to look at the spectral distribution $\diff I / \diff \Omega$ only in the redshift-receding limit $\theta \sim 0$, we immediately see the problem (as we have already pointed out).
The \textquote{temperature} of the uniformly accelerated electron in Eq.~\eqref{uni_temperature}
\begin{equation}
	T_{\textrm{UV}} = \frac{e^2\kappa}{4 \sin\theta }, \qquad \textrm{(Classical; Stoney Scale)}
\end{equation}
diverges at $\theta \to 0$.
This means that we cannot define the temperature for such an electron! Moreover, in the case of a mirror, the uniform acceleration non-thermality is consistent with the temporal-spatial observation that the uniformly accelerated trajectory does not have zero jerk; see Eq.~(\ref{jec13}).

\paragraph{Davies-Fulling electron is thermal}

With the Davies-Fulling electron, the situation is more straightforward. 
The temperature in the redshift-receding limit, Eq.~(\ref{T_red}), is perfectly finite.
Therefore, we can legitimately call it thermal.

\section{Conclusions}\label{sec:CONCLUSIONS}

Our approach was to investigate the thermodynamics of acceleration radiation for a moving point charge by exact mapping of canonical solutions associated with the moving mirror model. We have demonstrated the analytic utility of the electron-mirror correspondence and its ability to answer questions on the meaning of acceleration temperature in classical and quantum physical contexts.

Having explicitly confirmed the consistency of the electron-mirror duality, we summarize the main results of our approach: 
\begin{itemize}

    \item We have found the total energy,  Eq.~(\ref{totalenergy}) and Eq.~(\ref{wd_total_energy}), emitted by the Davies-Fulling electron, Eq.~(\ref{eom}), and the Walker-Davies electron, Eq.~(\ref{wd_traj}).  These expressions are critical for confirming consistency. They physically reveal that the faster electrons move, the more energy is radiated. 
    \item We have found the spectral distributions of the Davies-Fulling Eq.~(\ref{dIdOmega_3}), Walker-Davies Eq.~(\ref{WD_SD}), and uniform proper acceleration Eq.~(\ref{recipe_dIdOmega_from_mirror_uniform}) [and Appendix Eq.~(\ref{recipe_dIdOmega_from_mirror_uniform2})], electron worldlines. The Davies-Fulling electron manifests the Planck factor, Eq.~(\ref{DF_dIdOmega_redshift}).
\item In Eq.~(\ref{vmaxPARTICLE}), global particle count is proportional to the square of the maximum speed, $N = v_{\textrm{max}}^2/6$, for the Walker-Davies trajectory (non-relativistic regime).  This is the only example of analytic tractability for total particle creation for classical radiation emitted by a moving point charge for a globally defined non-periodic worldline.
    \item We have found the temperature, Eq.~(\ref{T_red}), of the radiation emitted by the Davies-Fulling electron. Interestingly, it is dependent on the final speed of the electron. The faster it goes, the hotter it glows.
   
    \item We have shown that zero jerk results in thermality.  This is the dynamic-thermal explanation for the temperature of an accelerating moving mirror, Eq.~(\ref{jerktemp}); explicitly shown in the case of the Davies-Fulling and Carlitz-Willey.
\end{itemize}
There are several limitations to our approach.  In particular, several specific drawbacks should be pointed out to avoid misinterpretation or misapplication of its results, potentially resulting in erroneous conclusions, namely: 
\begin{enumerate}
    \item None of the spectral distributions can be analytically integrated over their solid angle. This contrasts with the infrared acceleration trajectory associated with beta decay, i.e., \cite{Good:2022eub,Ievlev:2023inj,Lynch:2022rqx}. This frustrates attempts at finding the most general form of the frequency spectrum $I(\omega)$, the particle spectrum $N(\omega)$, or an analytic total particle count $N$. Without a known spectrum $I(\omega)$, the spectral-statistics (as explicitly derived from the
spectral distribution in a particular angular regime for
the radiation from a moving point charge) do not necessarily characterize the spin-statistics of the electromagnetic field, e.g., \cite{Ievlev:2023akh}.
    \item Since the Planck factor of the Davies-Fulling electron has only been found in the spectral distribution $\diff{I}/\diff{\Omega}$, we do not have an associated temperature for the spectrum $I(\omega)$ itself.  This is arguably a kind of quasi-thermality \cite{Good:2019tnf}, consistent with unitarity when $s<1$; applicable only in an angular regime rather than a cumulative global regime.
    \item While the approach of examining exact solutions has shed light on the application and demonstrated the consistency of the correspondence and the meaning of temperature across the electron-mirror contexts, the approach does not shed light on the physical origin of frequency-angular mapping itself; some effort in this direction is given in \cite{Ritus:2022bph}. 
    
\end{enumerate}

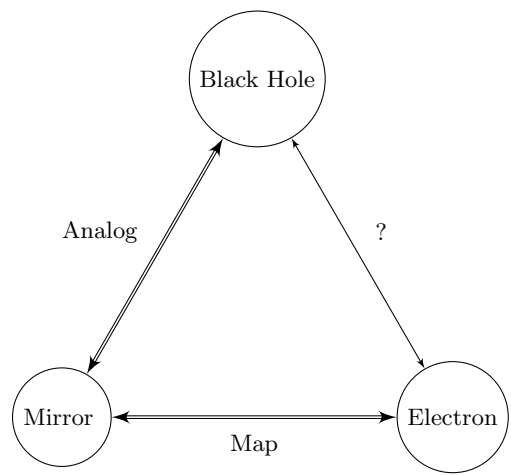
\begin{figure}[htbp]
\begin{center}
\begin{tikzpicture}
\node[draw,circle] (A) at (90:3) {Black Hole};
\node[draw,circle] (B) at (210:3) {Mirror\;};
\node[draw,circle] (C) at (330:3) {Electron};
\draw[latex'-latex',double] (A) -- node[label=150:Analog,label=330:] {} (B);
\draw[latex'-latex'] (A) -- node[label=30:?,label=210:] {} (C);
\draw[latex'-latex',double] (B) -- node[label=90:,label=270:Map] {} (C);
\end{tikzpicture}
\end{center}
\caption{A conceptual map demonstrating the context of the electron-mirror duality as the lower leg of a thermal-particle-creation triangle.  The mathematical identity map between the classical/quantum model of the electron and mirror is exact. Since the moving mirror has a half-century history as a useful analog for black hole radiation, the least developed side of the triangle is the link between experimentally-challenging black holes and experimentation-friendly electrons.  Our electron results explicitly demonstrate the duality's power, providing tractable means for investigating the classical quantum aspects of relativistic thermodynamics (a languishing field with a severe paucity of experimental evidence). Our solutions help to reorient the direction of inquiry and establish analytic examples that illustrate the `linographic' (physical content in 3D space is encoded in 1D space) utility of the electron-mirror duality.}  
\label{Concept_Map}
\end{figure}

Despite these limitations, the moving mirror model and the moving point charge share a functional mathematical identity (see Fig. \ref{Concept_Map} for a concept map) when their worldlines share dynamic invariance across different spatial dimensions. The distinct physical meaning of temperature within the two systems unveils insight into the fundamental nature of radiative thermodynamics.

\section{Acknowledgements} Funding comes partly from the FY2024-SGP-1-STMM Faculty Development Competitive Research Grant (FDCRGP) no.201223FD8824 and SSH20224004 at Nazarbayev University in Qazaqstan. Appreciation is given to the ROC (Taiwan) Ministry of Science and Technology (MOST), Grant no.112-2112-M-002-013, National Center for Theoretical Sciences (NCTS), and Leung Center for Cosmology and Particle Astrophysics (LeCosPA) of National Taiwan University.        

\appendix

\section{Spectral distribution for the uniformly proper accelerated charge} \label{Appendix:UA}

This section will show how to derive the spectral distribution Eq.~\eqref{recipe_dIdOmega_from_mirror_uniform} from electrodynamics without employing the electron-mirror duality.
To this end, we will use Eq.~\eqref{I_Omega_Jackson} and Eq.~\eqref{jz_fourier_transform_definition}.

Let us start with a simple case $\kappa = 1$ (we will recover $\kappa$ at the end).
The trajectory, in this case, is given by
\begin{equation}
	z(t) = 1 - \sqrt{1 + t^2} \,.
\end{equation}
Passing to the retarded time $u = t - z(t)$ brings the trajectory into the form
\begin{equation}
	t = \frac{ u (2 + u) }{ 2(1 + u) } \,,
\end{equation}
and the Fourier transform of the current Eq.~\eqref{jz_fourier_transform_definition} can be written as an integral over $u$:
\begin{equation}
	j_z (\omega, k_z)
		= - e\, \int\limits_{- 1}^{\infty} \diff{u} \, 
		\frac{ u (2 + u) }{ 2(1 + u)^2 } e^{ i \omega \frac{ u ( 2 + u + u \cos\theta ) }{ 2 (1+u) } } \,,
\end{equation}
where we substituted $k_z = \omega \cos \theta$.
Making a further change of variables $u = U-1$ we can bring the integral to the form of the modified Bessel function of the second
kind:
\begin{equation}
\begin{aligned}
	j_z& (\omega, k_z) \\
		&= - \frac{e}{2} \, e^{- i \omega \cos\theta } \int\limits_{0}^{\infty} \diff{U} \, 
			\left( 1 - \frac{1}{U^2} \right)
			e^{ \frac{ i \omega (\cos\theta + 1) }{ 2 } \left( U + \frac{ \cos\theta - 1 }{ \cos\theta + 1 } \frac{1}{U} \right) } \\
		&= - \frac{e}{2} \, e^{- i \omega \cos\theta } \Big[ 
				2 \sqrt{ \frac{\cos\theta - 1}{\cos\theta + 1} } K_{-1} (\omega \sin\theta ) \\
				&\phantom{===========} - 2 \sqrt{ \frac{\cos\theta + 1}{\cos\theta - 1} } K_{1}  (\omega \sin\theta )
			\Big] \\
		&= \frac{ 2 i e }{ \sin\theta } \, e^{- i \omega \cos\theta } K_{1}  (\omega \sin\theta ) \,.
\end{aligned}
\end{equation}
Here we have used Eq. (8.432.7)  
and Eq. (8.486.10) 
from the book of Gradshteyn and Ryzhik (7$^\text{th}$ ed., 2007).

Finally, using Eq.~\eqref{I_Omega_Jackson} we arrive at the spectral distribution:
\begin{equation}
    \frac{\diff{I}}{\diff{\Omega}} = \frac{e^2 \omega^2}{4\pi^3 } \left|K_1\left( \omega \sin\theta \right)\right|^2 \,, \quad
    \kappa = 1 \,.
\end{equation}
The acceleration parameter $\kappa$ can be easily restored by dimensional analysis, and we arrive at
\begin{equation}
    \frac{\diff{I}}{\diff{\Omega}} = \frac{e^2 \omega^2}{4\pi^3 \kappa^2 } \left|K_1\left(\frac{\omega }{\kappa} \sin\theta \right)\right|^2 \,.
\label{recipe_dIdOmega_from_mirror_uniform2}
\end{equation}
This coincides with the result Eq.~\eqref{recipe_dIdOmega_from_mirror_uniform} obtained from the electron-mirror duality.
\section{Zero jerk and thermality}\label{Appendix:JERK}
In this section, we will demonstrate the non-intuitive relativistic result that zero jerk, $J=0$, does not result in a constant proper acceleration but instead gives a proper acceleration which has an inverse proper time dependence, $\alpha(\tau)= \tau^{-1}$.  This inverse dependence has been derived in the literature \cite{Good:2017ddq}, and it belongs to the worldline of Carlitz-Willey, associated with eternal thermal radiation from a moving mirror \cite{carlitz1987reflections} or, if you will, an analog eternally evaporating black hole with temperature $T = \kappa/2\pi$.

First, we recall the spacetime vector, $X^\mu = (t,x)$, and its velocity, acceleration and jerk:
\be V^\mu = \dot{X}^\mu, \quad A^\mu = \ddot{X}^\mu, \quad J^\mu = \dddot{X}^\mu,\ee
where the dot implies a derivative with respect to proper time, $\tau$. These quantities can be written in terms of the Lorentz factor $\gamma = (1-v^2)^{-1/2}$, and the celerity $w = \gamma v$:
\be V^\mu = (\gamma,w), \quad A^\mu = (\dot{\gamma},\dot{w}), \quad J^\mu = (\ddot{\gamma},\ddot{w}),\ee
where the usual coordinate lab time velocity is $v = d x/dt$.
Using the rapidity $\eta$ and the proper acceleration $\alpha$, we write $\gamma = \cosh\eta$, $w = \sinh\eta$, and $\dot{\eta} = \alpha$, obtaining
\be A^\mu = (\dot{\gamma},\dot{w}) = (\dot{\eta}\sinh\eta ,\dot{\eta}\cosh\eta) = \alpha(w,\gamma).\ee 
The jerk vector is also easily found,
\begin{align}
J^\mu &= (\ddot{\gamma},\ddot{w}),\\ 
&= (\dot{\eta}^2 \cosh\eta+ \ddot{\eta}\sinh\eta,\dot{\eta}^2 \sinh\eta+ \ddot{\eta}\cosh\eta),\\
&= (\gamma \alpha^2 + w\dot{\alpha},w\alpha^2 + \gamma \dot{\alpha}).
\end{align}
Therefore, using $\gamma^2 - w^2 = 1$, the jerk modulus is 
\begin{equation}\label{jec13}
J^2 = J_\nu J^\nu=\alpha^4-\dot{\alpha}^2.
\end{equation}
The Lorentz invariant $J$ is called the proper jerk, in a similar way one calls the Lorentz invariant $\alpha$ the proper acceleration. 
In summary, the results are:
\begin{align}
V^\mu &= (\gamma,w),  \quad A^\mu = \alpha(w,\gamma),  \\ 
J^\mu &= (\gamma \alpha^2 + w\dot{\alpha},w\alpha^2 + \gamma \dot{\alpha}), 
\end{align}
and the corresponding Lorentz invariants
\begin{align}
 V_\mu V^\mu &=1, \quad A_\mu A^\mu  =-\alpha^2, \quad
J_\nu J^\nu  =\alpha^4-\dot{\alpha}^2.  
\end{align}
Note that $V_\mu A^\mu=0$, and $V_\mu J^\mu = \alpha^2$.  It is straightforward to verify that 
\be J^\mu = \frac{\dot{\alpha}}{\alpha}A^\mu + \alpha^2V^\mu.\ee
We can now examine the simplest case of motion when the jerk is absent.  

Let us look at the proper acceleration when the proper jerk is zero.  Writing a vanishing modulus jerk with two roots gives:
\be J^2 = (\alpha^2 + \dot{\alpha})(\alpha^2 - \dot{\alpha}) = 0,\label{modJ2}\ee
where we can choose an accelerated motion 
\be \alpha^2 + \dot{\alpha} = 0.\label{j0}\ee
This has the proper time-dependent solution,
\begin{equation}
    |\alpha(\tau)| =  \frac{1}{\tau}. 
\label{invtau}
\end{equation}
Again, recall the proper jerk has been set to zero in Eq.~(\ref{modJ2}). Therefore, Eq.~(\ref{invtau}) is the proper acceleration without proper jerk.  

This is non-intuitive because the proper jerk for uniform proper acceleration is not zero.  Consider $\alpha_0$ as a non-zero constant for eternal proper uniform acceleration.  Using Eq.~(\ref{modJ2}), $J^2 = \alpha^4 -\dot{\alpha}^2 \neq 0$, one obtains non-zero constant proper jerk, $J_0 = \pm \alpha_0^2$.  

The eternal thermal Carlitz-Willey trajectory \cite{carlitz1987reflections}, has proper acceleration given by Eq.~(\ref{invtau}), $\alpha(\tau) = \tau^{-1}$ \cite{Good:2017ddq}.  It is well-known to have a constant peel acceleration, $\mathcal{P} = 2\alpha e^{\eta} = \kappa$ \cite{CW2lifetime}.  Here $\mathcal{P}$ is the peel, and $\kappa$ is the acceleration parameter (a constant) associated with the temperature of the radiation $T = \kappa/2\pi$.  The $s=1$ Davies-Fulling mirror, Eq.~(\ref{eom}), $\alpha = -\kappa \gamma$ also has a constant peel at late coordinate times near $\tau_0 = \pi/2\kappa$ because the proper acceleration scales as $\alpha = 1/(\tau -\tau_0)$ \cite{Good:2017ddq}.  

Because thermal moving mirrors have constant peel acceleration (or $\sim 1/\tau$ proper acceleration), it is worthwhile to consider the proper time derivative of the peel: 
\be \mathcal{P} = 2\alpha e^{\eta},\quad \dot{\mathcal{P}} = 2 e^\eta \alpha^2  + 2 e^\eta \dot{\alpha}.\ee
If $\dot{\mathcal{P}} = 0$, then in agreement with Eq.~(\ref{j0}),
\be \dot{\mathcal{P}} = \alpha^2 + \dot{\alpha} = 0.\ee
Thus, one can see clearly that zero jerk is responsible for the temperature of the moving mirror:
\be J = 0 \quad \rightarrow \quad T = \frac{\hbar\kappa}{2\pi}. \label{jerktemp}\ee
This directly connects motion and thermal radiation in the moving mirror model.  No such connection has been found for thermal radiation emitted by a moving point charge. 



\bibliography{main} 

\begin{thebibliography}{62}%
\makeatletter
\providecommand \@ifxundefined [1]{%
 \@ifx{#1\undefined}
}%
\providecommand \@ifnum [1]{%
 \ifnum #1\expandafter \@firstoftwo
 \else \expandafter \@secondoftwo
 \fi
}%
\providecommand \@ifx [1]{%
 \ifx #1\expandafter \@firstoftwo
 \else \expandafter \@secondoftwo
 \fi
}%
\providecommand \natexlab [1]{#1}%
\providecommand \enquote  [1]{``#1''}%
\providecommand \bibnamefont  [1]{#1}%
\providecommand \bibfnamefont [1]{#1}%
\providecommand \citenamefont [1]{#1}%
\providecommand \href@noop [0]{\@secondoftwo}%
\providecommand \href [0]{\begingroup \@sanitize@url \@href}%
\providecommand \@href[1]{\@@startlink{#1}\@@href}%
\providecommand \@@href[1]{\endgroup#1\@@endlink}%
\providecommand \@sanitize@url [0]{\catcode `\\12\catcode `\$12\catcode
  `\&12\catcode `\#12\catcode `\^12\catcode `\_12\catcode `\%12\relax}%
\providecommand \@@startlink[1]{}%
\providecommand \@@endlink[0]{}%
\providecommand \url  [0]{\begingroup\@sanitize@url \@url }%
\providecommand \@url [1]{\endgroup\@href {#1}{\urlprefix }}%
\providecommand \urlprefix  [0]{URL }%
\providecommand \Eprint [0]{\href }%
\providecommand \doibase [0]{http://dx.doi.org/}%
\providecommand \selectlanguage [0]{\@gobble}%
\providecommand \bibinfo  [0]{\@secondoftwo}%
\providecommand \bibfield  [0]{\@secondoftwo}%
\providecommand \translation [1]{[#1]}%
\providecommand \BibitemOpen [0]{}%
\providecommand \bibitemStop [0]{}%
\providecommand \bibitemNoStop [0]{.\EOS\space}%
\providecommand \EOS [0]{\spacefactor3000\relax}%
\providecommand \BibitemShut  [1]{\csname bibitem#1\endcsname}%
\let\auto@bib@innerbib\@empty
\bibitem [{\citenamefont {Davies}(1975)}]{Davies:1974th}%
  \BibitemOpen
  \bibfield  {author} {\bibinfo {author} {\bibfnamefont {P.~C.~W.}\
  \bibnamefont {Davies}},\ }\bibfield  {title} {\enquote {\bibinfo {title}
  {{Scalar particle production in Schwarzschild and Rindler metrics}},}\ }\href
  {\doibase 10.1088/0305-4470/8/4/022} {\bibfield  {journal} {\bibinfo
  {journal} {J. Phys. A}\ }\textbf {\bibinfo {volume} {8}},\ \bibinfo {pages}
  {609--616} (\bibinfo {year} {1975})}\BibitemShut {NoStop}%
\bibitem [{\citenamefont {Unruh}(1976)}]{unruh76}%
  \BibitemOpen
  \bibfield  {author} {\bibinfo {author} {\bibfnamefont {W.~G.}\ \bibnamefont
  {Unruh}},\ }\bibfield  {title} {\enquote {\bibinfo {title} {Notes on
  black-hole evaporation},}\ }\href {\doibase 10.1103/PhysRevD.14.870}
  {\bibfield  {journal} {\bibinfo  {journal} {Phys. Rev. D}\ }\textbf {\bibinfo
  {volume} {14}},\ \bibinfo {pages} {870--892} (\bibinfo {year}
  {1976})}\BibitemShut {NoStop}%
\bibitem [{\citenamefont {Hawking}(1974)}]{Hawking:1974rv}%
  \BibitemOpen
  \bibfield  {author} {\bibinfo {author} {\bibfnamefont {S.~W.}\ \bibnamefont
  {Hawking}},\ }\bibfield  {title} {\enquote {\bibinfo {title} {{Black hole
  explosions}},}\ }\href {\doibase 10.1038/248030a0} {\bibfield  {journal}
  {\bibinfo  {journal} {Nature}\ }\textbf {\bibinfo {volume} {248}},\ \bibinfo
  {pages} {30--31} (\bibinfo {year} {1974})}\BibitemShut {NoStop}%
\bibitem [{\citenamefont {Hawking}(1975)}]{Hawking:1974sw}%
  \BibitemOpen
  \bibfield  {author} {\bibinfo {author} {\bibfnamefont {S.W.}\ \bibnamefont
  {Hawking}},\ }\bibfield  {title} {\enquote {\bibinfo {title} {{Particle
  Creation by Black Holes}},}\ }\href {\doibase 10.1007/BF02345020} {\bibfield
  {journal} {\bibinfo  {journal} {Commun. Math. Phys.}\ }\textbf {\bibinfo
  {volume} {43}},\ \bibinfo {pages} {199--220} (\bibinfo {year}
  {1975})}\BibitemShut {NoStop}%
\bibitem [{\citenamefont {Penrose}(1996)}]{Penrose:1996cv}%
  \BibitemOpen
  \bibfield  {author} {\bibinfo {author} {\bibfnamefont {R.}~\bibnamefont
  {Penrose}},\ }\bibfield  {title} {\enquote {\bibinfo {title} {{On gravity's
  role in quantum state reduction}},}\ }\href {\doibase 10.1007/BF02105068}
  {\bibfield  {journal} {\bibinfo  {journal} {Gen. Rel. Grav.}\ }\textbf
  {\bibinfo {volume} {28}},\ \bibinfo {pages} {581--600} (\bibinfo {year}
  {1996})}\BibitemShut {NoStop}%
\bibitem [{\citenamefont {DeWitt}(1980)}]{dewitt_detector}%
  \BibitemOpen
  \bibfield  {author} {\bibinfo {author} {\bibfnamefont {B.}~\bibnamefont
  {DeWitt}},\ }\href@noop {} {\emph {\bibinfo {title} {{General Relativity}:
  {An Einstein Centenary Survey}}}}\ (\bibinfo  {publisher} {Univ. Pr.},\
  \bibinfo {address} {Cambridge, UK},\ \bibinfo {year} {1980})\BibitemShut
  {NoStop}%
\bibitem [{\citenamefont {Moore}(1970)}]{moore1970quantum}%
  \BibitemOpen
  \bibfield  {author} {\bibinfo {author} {\bibfnamefont {Gerald~T.}\
  \bibnamefont {Moore}},\ }\bibfield  {title} {\enquote {\bibinfo {title}
  {Quantum theory of the electromagnetic field in a variable‐length
  one‐dimensional cavity},}\ }\href {https://doi.org/10.1063/1.1665432}
  {\bibfield  {journal} {\bibinfo  {journal} {J. of Math. Phys.}\ }\textbf
  {\bibinfo {volume} {11}},\ \bibinfo {pages} {2679--2691} (\bibinfo {year}
  {1970})}\BibitemShut {NoStop}%
\bibitem [{\citenamefont {Fulling}(1973)}]{Fulling:1972md}%
  \BibitemOpen
  \bibfield  {author} {\bibinfo {author} {\bibfnamefont {Stephen~A.}\
  \bibnamefont {Fulling}},\ }\bibfield  {title} {\enquote {\bibinfo {title}
  {{Nonuniqueness of canonical field quantization in Riemannian space-time}},}\
  }\href {\doibase 10.1103/PhysRevD.7.2850} {\bibfield  {journal} {\bibinfo
  {journal} {Phys. Rev. D}\ }\textbf {\bibinfo {volume} {7}},\ \bibinfo {pages}
  {2850--2862} (\bibinfo {year} {1973})}\BibitemShut {NoStop}%
\bibitem [{\citenamefont {Fulling}\ and\ \citenamefont
  {Davies}(1976)}]{Davies:1976hi}%
  \BibitemOpen
  \bibfield  {author} {\bibinfo {author} {\bibfnamefont {S.~A.}\ \bibnamefont
  {Fulling}}\ and\ \bibinfo {author} {\bibfnamefont {P.~C.~W.}\ \bibnamefont
  {Davies}},\ }\bibfield  {title} {\enquote {\bibinfo {title} {Radiation from a
  moving mirror in two dimensional space-time: conformal anomaly},}\ }\href
  {https://royalsocietypublishing.org/doi/abs/10.1098/rspa.1976.0045}
  {\bibfield  {journal} {\bibinfo  {journal} {Proc. R. Soc. Lond. A}\ }\textbf
  {\bibinfo {volume} {348}},\ \bibinfo {pages} {393--414} (\bibinfo {year}
  {1976})}\BibitemShut {NoStop}%
\bibitem [{\citenamefont {Davies}\ and\ \citenamefont
  {Fulling}(1977)}]{Davies:1977yv}%
  \BibitemOpen
  \bibfield  {author} {\bibinfo {author} {\bibfnamefont {P.C.W.}\ \bibnamefont
  {Davies}}\ and\ \bibinfo {author} {\bibfnamefont {S.A.}\ \bibnamefont
  {Fulling}},\ }\bibfield  {title} {\enquote {\bibinfo {title} {{Radiation from
  Moving Mirrors and from Black Holes}},}\ }\href {\doibase
  10.1098/rspa.1977.0130} {\bibfield  {journal} {\bibinfo  {journal} {Proc. R.
  Soc. Lond. A}\ }\textbf {\bibinfo {volume} {A356}},\ \bibinfo {pages}
  {237--257} (\bibinfo {year} {1977})}\BibitemShut {NoStop}%
\bibitem [{\citenamefont {Unruh}\ and\ \citenamefont
  {Wald}(1982)}]{Unruh:1982ic}%
  \BibitemOpen
  \bibfield  {author} {\bibinfo {author} {\bibfnamefont {W.~G.}\ \bibnamefont
  {Unruh}}\ and\ \bibinfo {author} {\bibfnamefont {Robert~M.}\ \bibnamefont
  {Wald}},\ }\bibfield  {title} {\enquote {\bibinfo {title} {{Acceleration
  Radiation and Generalized Second Law of Thermodynamics}},}\ }\href {\doibase
  10.1103/PhysRevD.25.942} {\bibfield  {journal} {\bibinfo  {journal} {Phys.
  Rev. D}\ }\textbf {\bibinfo {volume} {25}},\ \bibinfo {pages} {942--958}
  (\bibinfo {year} {1982})}\BibitemShut {NoStop}%
\bibitem [{\citenamefont {Ford}\ and\ \citenamefont
  {Vilenkin}(1982)}]{Ford:1982ct}%
  \BibitemOpen
  \bibfield  {author} {\bibinfo {author} {\bibfnamefont {L.H.}\ \bibnamefont
  {Ford}}\ and\ \bibinfo {author} {\bibfnamefont {Alexander}\ \bibnamefont
  {Vilenkin}},\ }\bibfield  {title} {\enquote {\bibinfo {title} {{Quantum
  radiation by moving mirrors}},}\ }\href {\doibase 10.1103/PhysRevD.25.2569}
  {\bibfield  {journal} {\bibinfo  {journal} {Phys. Rev. D}\ }\textbf {\bibinfo
  {volume} {25}},\ \bibinfo {pages} {2569} (\bibinfo {year}
  {1982})}\BibitemShut {NoStop}%
\bibitem [{\citenamefont {Nikishov}\ and\ \citenamefont
  {Ritus}(1995)}]{Nikishov:1995qs}%
  \BibitemOpen
  \bibfield  {author} {\bibinfo {author} {\bibfnamefont {A.I.}\ \bibnamefont
  {Nikishov}}\ and\ \bibinfo {author} {\bibfnamefont {V.I.}\ \bibnamefont
  {Ritus}},\ }\bibfield  {title} {\enquote {\bibinfo {title} {{Emission of
  scalar photons by an accelerated mirror in (1+1) space and its relation to
  the radiation from an electrical charge in classical electrodynamics}},}\
  }\href@noop {} {\bibfield  {journal} {\bibinfo  {journal} {J. Exp. Theor.
  Phys.}\ }\textbf {\bibinfo {volume} {81}},\ \bibinfo {pages} {615--624}
  (\bibinfo {year} {1995})}\BibitemShut {NoStop}%
\bibitem [{\citenamefont {Chen}\ \emph {et~al.}(2022)\citenamefont {Chen} \emph
  {et~al.}}]{AnaBHEL:2022sri}%
  \BibitemOpen
  \bibfield  {author} {\bibinfo {author} {\bibfnamefont {Pisin}\ \bibnamefont
  {Chen}} \emph {et~al.} (\bibinfo {collaboration} {AnaBHEL}),\ }\bibfield
  {title} {\enquote {\bibinfo {title} {{AnaBHEL (Analog Black Hole Evaporation
  via Lasers) Experiment: Concept, Design, and Status}},}\ }\href {\doibase
  10.3390/photonics9121003} {\bibfield  {journal} {\bibinfo  {journal}
  {Photon.}\ }\textbf {\bibinfo {volume} {9}},\ \bibinfo {pages} {1003}
  (\bibinfo {year} {2022})},\ \Eprint {http://arxiv.org/abs/2205.12195}
  {arXiv:2205.12195 [gr-qc]} \BibitemShut {NoStop}%
\bibitem [{\citenamefont {Chen}\ and\ \citenamefont
  {Mourou}(2017)}]{Chen:2015bcg}%
  \BibitemOpen
  \bibfield  {author} {\bibinfo {author} {\bibfnamefont {Pisin}\ \bibnamefont
  {Chen}}\ and\ \bibinfo {author} {\bibfnamefont {Gerard}\ \bibnamefont
  {Mourou}},\ }\bibfield  {title} {\enquote {\bibinfo {title} {{Accelerating
  Plasma Mirrors to Investigate Black Hole Information Loss Paradox}},}\ }\href
  {\doibase 10.1103/PhysRevLett.118.045001} {\bibfield  {journal} {\bibinfo
  {journal} {Phys. Rev. Lett.}\ }\textbf {\bibinfo {volume} {118}},\ \bibinfo
  {pages} {045001} (\bibinfo {year} {2017})},\ \Eprint
  {http://arxiv.org/abs/1512.04064} {arXiv:1512.04064 [gr-qc]} \BibitemShut
  {NoStop}%
\bibitem [{\citenamefont {Chen}\ and\ \citenamefont
  {Mourou}(2020)}]{Chen:2020sir}%
  \BibitemOpen
  \bibfield  {author} {\bibinfo {author} {\bibfnamefont {Pisin}\ \bibnamefont
  {Chen}}\ and\ \bibinfo {author} {\bibfnamefont {Gerard}\ \bibnamefont
  {Mourou}},\ }\bibfield  {title} {\enquote {\bibinfo {title} {{Trajectory of a
  flying plasma mirror traversing a target with density gradient}},}\ }\href
  {\doibase 10.1063/5.0012374} {\bibfield  {journal} {\bibinfo  {journal}
  {Phys. Plasmas}\ }\textbf {\bibinfo {volume} {27}},\ \bibinfo {pages}
  {123106} (\bibinfo {year} {2020})},\ \Eprint
  {http://arxiv.org/abs/2004.10615} {arXiv:2004.10615 [physics.plasm-ph]}
  \BibitemShut {NoStop}%
\bibitem [{\citenamefont {Steinhauer}(2014)}]{Steinhauer:2014dra}%
  \BibitemOpen
  \bibfield  {author} {\bibinfo {author} {\bibfnamefont {Jeff}\ \bibnamefont
  {Steinhauer}},\ }\bibfield  {title} {\enquote {\bibinfo {title} {{Observation
  of self-amplifying Hawking radiation in an analog black hole laser}},}\
  }\href {\doibase 10.1038/NPHYS3104} {\bibfield  {journal} {\bibinfo
  {journal} {Nature Phys.}\ }\textbf {\bibinfo {volume} {10}},\ \bibinfo
  {pages} {864} (\bibinfo {year} {2014})},\ \Eprint
  {http://arxiv.org/abs/1409.6550} {arXiv:1409.6550 [cond-mat.quant-gas]}
  \BibitemShut {NoStop}%
\bibitem [{\citenamefont {Lynch}\ \emph {et~al.}(2023)\citenamefont {Lynch},
  \citenamefont {Ievlev},\ and\ \citenamefont {Good}}]{Lynch:2022rqx}%
  \BibitemOpen
  \bibfield  {author} {\bibinfo {author} {\bibfnamefont {Morgan~H}\
  \bibnamefont {Lynch}}, \bibinfo {author} {\bibfnamefont {Evgenii}\
  \bibnamefont {Ievlev}}, \ and\ \bibinfo {author} {\bibfnamefont {Michael
  R~R}\ \bibnamefont {Good}},\ }\bibfield  {title} {\enquote {\bibinfo {title}
  {{Accelerated electron thermometer: observation of 1D Planck radiation}},}\
  }\href {\doibase 10.1093/ptep/ptad157} {\bibfield  {journal} {\bibinfo
  {journal} {Progress of Theoretical and Experimental Physics}\ ,\ \bibinfo
  {pages} {ptad157}} (\bibinfo {year} {2023})},\ \Eprint
  {http://arxiv.org/abs/2211.14774} {arXiv:2211.14774 [nucl-ex]} \BibitemShut
  {NoStop}%
\bibitem [{\citenamefont {Bales}\ \emph {et~al.}(2016)\citenamefont {Bales}
  \emph {et~al.}}]{RDKII:2016lpd}%
  \BibitemOpen
  \bibfield  {author} {\bibinfo {author} {\bibfnamefont {M.~J.}\ \bibnamefont
  {Bales}} \emph {et~al.} (\bibinfo {collaboration} {RDK II}),\ }\bibfield
  {title} {\enquote {\bibinfo {title} {{Precision Measurement of the Radiative
  $\beta$ Decay of the Free Neutron}},}\ }\href {\doibase
  10.1103/PhysRevLett.116.242501} {\bibfield  {journal} {\bibinfo  {journal}
  {Phys. Rev. Lett.}\ }\textbf {\bibinfo {volume} {116}},\ \bibinfo {pages}
  {242501} (\bibinfo {year} {2016})},\ \Eprint
  {http://arxiv.org/abs/1603.00243} {arXiv:1603.00243 [nucl-ex]} \BibitemShut
  {NoStop}%
\bibitem [{\citenamefont {Nico}\ \emph {et~al.}(2006)\citenamefont {Nico} \emph
  {et~al.}}]{nico}%
  \BibitemOpen
  \bibfield  {author} {\bibinfo {author} {\bibfnamefont {J.S.}\ \bibnamefont
  {Nico}} \emph {et~al.},\ }\bibfield  {title} {\enquote {\bibinfo {title}
  {{Observation of the radioactive decay mode of the free neutron}},}\ }\href
  {\doibase 10.1038/nature05390} {\bibfield  {journal} {\bibinfo  {journal}
  {Nature}\ }\textbf {\bibinfo {volume} {444}},\ \bibinfo {pages} {1059}
  (\bibinfo {year} {2006})}\BibitemShut {NoStop}%
\bibitem [{\citenamefont {F.R.S.}(1897)}]{Larmor1897}%
  \BibitemOpen
  \bibfield  {author} {\bibinfo {author} {\bibfnamefont {J.~Larmor~D.Sc.}\
  \bibnamefont {F.R.S.}},\ }\bibfield  {title} {\enquote {\bibinfo {title}
  {Lxiii. on the theory of the magnetic influence on spectra; and on the
  radiation from moving ions},}\ }\href {\doibase 10.1080/14786449708621095}
  {\bibfield  {journal} {\bibinfo  {journal} {The London, Edinburgh, and Dublin
  Philosophical Magazine and Journal of Science}\ }\textbf {\bibinfo {volume}
  {44}},\ \bibinfo {pages} {503--512} (\bibinfo {year} {1897})}\BibitemShut
  {NoStop}%
\bibitem [{\citenamefont {DeWitt}(1975)}]{DeWitt:1975ys}%
  \BibitemOpen
  \bibfield  {author} {\bibinfo {author} {\bibfnamefont {Bryce~S.}\
  \bibnamefont {DeWitt}},\ }\bibfield  {title} {\enquote {\bibinfo {title}
  {{Quantum Field Theory in Curved Space-Time}},}\ }\href {\doibase
  10.1016/0370-1573(75)90051-4} {\bibfield  {journal} {\bibinfo  {journal}
  {Phys. Rept.}\ }\textbf {\bibinfo {volume} {19}},\ \bibinfo {pages}
  {295--357} (\bibinfo {year} {1975})}\BibitemShut {NoStop}%
\bibitem [{\citenamefont {Ritus}(1998)}]{Ritus:1999eu}%
  \BibitemOpen
  \bibfield  {author} {\bibinfo {author} {\bibfnamefont {V.I.}\ \bibnamefont
  {Ritus}},\ }\bibfield  {title} {\enquote {\bibinfo {title} {{Symmetries and
  causes of the coincidence of the radiation spectra of mirrors and charges in
  (1+1) and (3+1) spaces}},}\ }\href {\doibase 10.1134/1.558646} {\bibfield
  {journal} {\bibinfo  {journal} {J. Exp. Theor. Phys.}\ }\textbf {\bibinfo
  {volume} {87}},\ \bibinfo {pages} {25--34} (\bibinfo {year} {1998})},\
  \Eprint {http://arxiv.org/abs/hep-th/9903083} {arXiv:hep-th/9903083}
  \BibitemShut {NoStop}%
\bibitem [{\citenamefont {Ritus}(2002)}]{Ritus:2002rq}%
  \BibitemOpen
  \bibfield  {author} {\bibinfo {author} {\bibfnamefont {V.I.}\ \bibnamefont
  {Ritus}},\ }\bibfield  {title} {\enquote {\bibinfo {title} {{Vacuum-vacuum
  amplitudes in the theory of quantum radiation by mirrors in 1+1-space and
  charges in 3+1-space}},}\ }\href {\doibase 10.1142/S0217751X02010467}
  {\bibfield  {journal} {\bibinfo  {journal} {Int. J. Mod. Phys. A}\ }\textbf
  {\bibinfo {volume} {17}},\ \bibinfo {pages} {1033--1040} (\bibinfo {year}
  {2002})}\BibitemShut {NoStop}%
\bibitem [{\citenamefont {Ritus}(2003)}]{Ritus:2003wu}%
  \BibitemOpen
  \bibfield  {author} {\bibinfo {author} {\bibfnamefont {V.I.}\ \bibnamefont
  {Ritus}},\ }\bibfield  {title} {\enquote {\bibinfo {title} {{The Symmetry,
  inferable from Bogoliubov transformation, between the processes induced by
  the mirror in two-dimensional and the charge in four-dimensional
  space-time}},}\ }\href {\doibase 10.1134/1.1600792} {\bibfield  {journal}
  {\bibinfo  {journal} {J. Exp. Theor. Phys.}\ }\textbf {\bibinfo {volume}
  {97}},\ \bibinfo {pages} {10--23} (\bibinfo {year} {2003})},\ \Eprint
  {http://arxiv.org/abs/hep-th/0309181} {arXiv:hep-th/0309181} \BibitemShut
  {NoStop}%
\bibitem [{\citenamefont {Zhakenuly}\ \emph {et~al.}(2021)\citenamefont
  {Zhakenuly}, \citenamefont {Temirkhan}, \citenamefont {Good},\ and\
  \citenamefont {Chen}}]{Zhakenuly:2021pfm}%
  \BibitemOpen
  \bibfield  {author} {\bibinfo {author} {\bibfnamefont {Abay}\ \bibnamefont
  {Zhakenuly}}, \bibinfo {author} {\bibfnamefont {Maksat}\ \bibnamefont
  {Temirkhan}}, \bibinfo {author} {\bibfnamefont {Michael R.~R.}\ \bibnamefont
  {Good}}, \ and\ \bibinfo {author} {\bibfnamefont {Pisin}\ \bibnamefont
  {Chen}},\ }\bibfield  {title} {\enquote {\bibinfo {title} {{Quantum power
  distribution of relativistic acceleration radiation: classical electrodynamic
  analogies with perfectly reflecting moving mirrors}},}\ }\href {\doibase
  10.3390/sym13040653} {\bibfield  {journal} {\bibinfo  {journal} {Symmetry}\
  }\textbf {\bibinfo {volume} {13}},\ \bibinfo {pages} {653} (\bibinfo {year}
  {2021})},\ \Eprint {http://arxiv.org/abs/2101.02511} {arXiv:2101.02511
  [gr-qc]} \BibitemShut {NoStop}%
\bibitem [{\citenamefont {Ritus}(2022)}]{Ritus:2022bph}%
  \BibitemOpen
  \bibfield  {author} {\bibinfo {author} {\bibfnamefont {V.~I.}\ \bibnamefont
  {Ritus}},\ }\bibfield  {title} {\enquote {\bibinfo {title} {{Finite value of
  the bare charge and the relation of the fine structure constant ratio for
  physical and bare charges to zero-point oscillations of the electromagnetic
  field in the vacuum}},}\ }\href {\doibase 10.3367/UFNe.2022.02.039167}
  {\bibfield  {journal} {\bibinfo  {journal} {Usp. Fiz. Nauk}\ }\textbf
  {\bibinfo {volume} {192}},\ \bibinfo {pages} {507--526} (\bibinfo {year}
  {2022})}\BibitemShut {NoStop}%
\bibitem [{\citenamefont {Ievlev}\ and\ \citenamefont
  {Good}(2024)}]{Ievlev:2023inj}%
  \BibitemOpen
  \bibfield  {author} {\bibinfo {author} {\bibfnamefont {Evgenii}\ \bibnamefont
  {Ievlev}}\ and\ \bibinfo {author} {\bibfnamefont {Michael R~R}\ \bibnamefont
  {Good}},\ }\bibfield  {title} {\enquote {\bibinfo {title} {{Thermal Larmor
  Radiation}},}\ }\href {\doibase 10.1093/ptep/ptae042} {\bibfield  {journal}
  {\bibinfo  {journal} {Progress of Theoretical and Experimental Physics}\ ,\
  \bibinfo {pages} {ptae042}} (\bibinfo {year} {2024})},\ \Eprint
  {http://arxiv.org/abs/2303.03676} {arXiv:2303.03676 [gr-qc]} \BibitemShut
  {NoStop}%
\bibitem [{\citenamefont {Ievlev}\ \emph {et~al.}(2024)\citenamefont {Ievlev},
  \citenamefont {Good},\ and\ \citenamefont {Linder}}]{Ievlev:2023bzk}%
  \BibitemOpen
  \bibfield  {author} {\bibinfo {author} {\bibfnamefont {Evgenii}\ \bibnamefont
  {Ievlev}}, \bibinfo {author} {\bibfnamefont {Michael~R.R.}\ \bibnamefont
  {Good}}, \ and\ \bibinfo {author} {\bibfnamefont {Eric~V.}\ \bibnamefont
  {Linder}},\ }\bibfield  {title} {\enquote {\bibinfo {title} {Ir-finite
  thermal acceleration radiation},}\ }\href {\doibase
  https://doi.org/10.1016/j.aop.2024.169593} {\bibfield  {journal} {\bibinfo
  {journal} {Annals of Physics}\ }\textbf {\bibinfo {volume} {461}},\ \bibinfo
  {pages} {169593} (\bibinfo {year} {2024})}\BibitemShut {NoStop}%
\bibitem [{\citenamefont {Good}\ and\ \citenamefont
  {Davies}(2023)}]{Good:2022eub}%
  \BibitemOpen
  \bibfield  {author} {\bibinfo {author} {\bibfnamefont {Michael R.~R.}\
  \bibnamefont {Good}}\ and\ \bibinfo {author} {\bibfnamefont {Paul C.~W.}\
  \bibnamefont {Davies}},\ }\bibfield  {title} {\enquote {\bibinfo {title}
  {{Infrared acceleration radiation}},}\ }\href {\doibase
  10.1007/s10701-023-00694-x} {\bibfield  {journal} {\bibinfo  {journal}
  {Foundations of Physics}\ }\textbf {\bibinfo {volume} {53}},\ \bibinfo
  {pages} {53} (\bibinfo {year} {2023})},\ \Eprint
  {http://arxiv.org/abs/2206.07291} {arXiv:2206.07291 [gr-qc]} \BibitemShut
  {NoStop}%
\bibitem [{\citenamefont {Good}(2023)}]{Good:2022xin}%
  \BibitemOpen
  \bibfield  {author} {\bibinfo {author} {\bibfnamefont {Michael R.~R.}\
  \bibnamefont {Good}},\ }\bibfield  {title} {\enquote {\bibinfo {title} {{On
  the temperature of lowest order inner bremsstrahlung}},}\ }\href@noop {} {\
  (\bibinfo {year} {2023})},\ \Eprint {http://arxiv.org/abs/2211.00946}
  {arXiv:2211.00946 [gr-qc]} \BibitemShut {NoStop}%
\bibitem [{\citenamefont {Good}\ \emph {et~al.}(2013)\citenamefont {Good},
  \citenamefont {Anderson},\ and\ \citenamefont {Evans}}]{good2013time}%
  \BibitemOpen
  \bibfield  {author} {\bibinfo {author} {\bibfnamefont {Michael R.~R.}\
  \bibnamefont {Good}}, \bibinfo {author} {\bibfnamefont {Paul~R.}\
  \bibnamefont {Anderson}}, \ and\ \bibinfo {author} {\bibfnamefont
  {Charles~R.}\ \bibnamefont {Evans}},\ }\bibfield  {title} {\enquote {\bibinfo
  {title} {Time dependence of particle creation from accelerating mirrors},}\
  }\href {\doibase 10.1103/PhysRevD.88.025023} {\bibfield  {journal} {\bibinfo
  {journal} {Phys. Rev. D}\ }\textbf {\bibinfo {volume} {88}},\ \bibinfo
  {pages} {025023} (\bibinfo {year} {2013})},\ \Eprint
  {http://arxiv.org/abs/1303.6756} {arXiv:1303.6756 [gr-qc]} \BibitemShut
  {NoStop}%
\bibitem [{\citenamefont {Birrell}\ and\ \citenamefont
  {Davies}(1984)}]{Birrell:1982ix}%
  \BibitemOpen
  \bibfield  {author} {\bibinfo {author} {\bibfnamefont {N.D.}\ \bibnamefont
  {Birrell}}\ and\ \bibinfo {author} {\bibfnamefont {P.C.W.}\ \bibnamefont
  {Davies}},\ }\href {\doibase 10.1017/CBO9780511622632} {\emph {\bibinfo
  {title} {{Quantum Fields in Curved Space}}}},\ Cambridge Monographs on
  Mathematical Physics\ (\bibinfo  {publisher} {Cambridge Univ. Press},\
  \bibinfo {address} {Cambridge, UK},\ \bibinfo {year} {1984})\BibitemShut
  {NoStop}%
\bibitem [{\citenamefont {Parker}\ and\ \citenamefont
  {Toms}(2009)}]{Parker:2009uva}%
  \BibitemOpen
  \bibfield  {author} {\bibinfo {author} {\bibfnamefont {Leonard~E.}\
  \bibnamefont {Parker}}\ and\ \bibinfo {author} {\bibfnamefont
  {D.}~\bibnamefont {Toms}},\ }\href {\doibase 10.1017/CBO9780511813924} {\emph
  {\bibinfo {title} {{Quantum Field Theory in Curved Spacetime}: {Quantized
  Field and Gravity}}}},\ Cambridge Monographs on Mathematical Physics\
  (\bibinfo  {publisher} {Cambridge University Press},\ \bibinfo {year}
  {2009})\BibitemShut {NoStop}%
\bibitem [{\citenamefont {Holzhey}\ \emph {et~al.}(1994)\citenamefont
  {Holzhey}, \citenamefont {Larsen},\ and\ \citenamefont
  {Wilczek}}]{Holzhey:1994we}%
  \BibitemOpen
  \bibfield  {author} {\bibinfo {author} {\bibfnamefont {Christoph}\
  \bibnamefont {Holzhey}}, \bibinfo {author} {\bibfnamefont {Finn}\
  \bibnamefont {Larsen}}, \ and\ \bibinfo {author} {\bibfnamefont {Frank}\
  \bibnamefont {Wilczek}},\ }\bibfield  {title} {\enquote {\bibinfo {title}
  {{Geometric and renormalized entropy in conformal field theory}},}\ }\href
  {\doibase 10.1016/0550-3213(94)90402-2} {\bibfield  {journal} {\bibinfo
  {journal} {Nucl. Phys. B}\ }\textbf {\bibinfo {volume} {424}},\ \bibinfo
  {pages} {443--467} (\bibinfo {year} {1994})},\ \Eprint
  {http://arxiv.org/abs/hep-th/9403108} {arXiv:hep-th/9403108} \BibitemShut
  {NoStop}%
\bibitem [{\citenamefont {Wilczek}(1993)}]{wilczek1993quantum}%
  \BibitemOpen
  \bibfield  {author} {\bibinfo {author} {\bibfnamefont {Frank}\ \bibnamefont
  {Wilczek}},\ }\bibfield  {title} {\enquote {\bibinfo {title} {{Quantum purity
  at a small price: Easing a black hole paradox}},}\ }in\ \href@noop {} {\emph
  {\bibinfo {booktitle} {{International Symposium on Black holes, Membranes,
  Wormholes and Superstrings}}}}\ (\bibinfo {year} {1993})\ pp.\ \bibinfo
  {pages} {1--21},\ \Eprint {http://arxiv.org/abs/hep-th/9302096}
  {arXiv:hep-th/9302096} \BibitemShut {NoStop}%
\bibitem [{\citenamefont {Walker}(1985{\natexlab{a}})}]{walker1985particle}%
  \BibitemOpen
  \bibfield  {author} {\bibinfo {author} {\bibfnamefont {W.~R.}\ \bibnamefont
  {Walker}},\ }\bibfield  {title} {\enquote {\bibinfo {title} {Particle and
  energy creation by moving mirrors},}\ }\href {\doibase
  10.1103/PhysRevD.31.767} {\bibfield  {journal} {\bibinfo  {journal} {Phys.
  Rev. D}\ }\textbf {\bibinfo {volume} {31}},\ \bibinfo {pages} {767--774}
  (\bibinfo {year} {1985}{\natexlab{a}})}\BibitemShut {NoStop}%
\bibitem [{\citenamefont {Walker}(1985{\natexlab{b}})}]{Walker:1984ya}%
  \BibitemOpen
  \bibfield  {author} {\bibinfo {author} {\bibfnamefont {W.R.}\ \bibnamefont
  {Walker}},\ }\bibfield  {title} {\enquote {\bibinfo {title} {{Negative Energy
  Fluxes and Moving Mirrors in Curved Space}},}\ }\href {\doibase
  10.1088/0264-9381/2/2/006} {\bibfield  {journal} {\bibinfo  {journal} {Class.
  Quant. Grav.}\ }\textbf {\bibinfo {volume} {2}},\ \bibinfo {pages} {L37}
  (\bibinfo {year} {1985}{\natexlab{b}})}\BibitemShut {NoStop}%
\bibitem [{\citenamefont {Carlitz}\ and\ \citenamefont
  {Willey}(1987{\natexlab{a}})}]{CW2lifetime}%
  \BibitemOpen
  \bibfield  {author} {\bibinfo {author} {\bibfnamefont {Robert~D.}\
  \bibnamefont {Carlitz}}\ and\ \bibinfo {author} {\bibfnamefont {Raymond~S.}\
  \bibnamefont {Willey}},\ }\bibfield  {title} {\enquote {\bibinfo {title}
  {Lifetime of a black hole},}\ }\href {\doibase 10.1103/PhysRevD.36.2336}
  {\bibfield  {journal} {\bibinfo  {journal} {Phys. Rev. D}\ }\textbf {\bibinfo
  {volume} {36}},\ \bibinfo {pages} {2336--2341} (\bibinfo {year}
  {1987}{\natexlab{a}})}\BibitemShut {NoStop}%
\bibitem [{\citenamefont {Ievlev}(2023)}]{Ievlev:2023ejs}%
  \BibitemOpen
  \bibfield  {author} {\bibinfo {author} {\bibfnamefont {Evgenii}\ \bibnamefont
  {Ievlev}},\ }\bibfield  {title} {\enquote {\bibinfo {title} {{Moving mirrors
  and event horizons in asymptotically non-flat spacetimes}},}\ }\href@noop {}
  {\  (\bibinfo {year} {2023})},\ \Eprint {http://arxiv.org/abs/2311.07403}
  {arXiv:2311.07403 [gr-qc]} \BibitemShut {NoStop}%
\bibitem [{\citenamefont {Osawa}\ \emph {et~al.}(2024)\citenamefont {Osawa},
  \citenamefont {Lin}, \citenamefont {Nambu}, \citenamefont {Hotta},\ and\
  \citenamefont {Chen}}]{Osawa:2024fqb}%
  \BibitemOpen
  \bibfield  {author} {\bibinfo {author} {\bibfnamefont {Yuki}\ \bibnamefont
  {Osawa}}, \bibinfo {author} {\bibfnamefont {Kuan-Nan}\ \bibnamefont {Lin}},
  \bibinfo {author} {\bibfnamefont {Yasusada}\ \bibnamefont {Nambu}}, \bibinfo
  {author} {\bibfnamefont {Masahiro}\ \bibnamefont {Hotta}}, \ and\ \bibinfo
  {author} {\bibfnamefont {Pisin}\ \bibnamefont {Chen}},\ }\bibfield  {title}
  {\enquote {\bibinfo {title} {{The final burst of the moving mirror is
  unrelated to the partner mode of analog Hawking radiation}},}\ }\href@noop {}
  {\  (\bibinfo {year} {2024})},\ \Eprint {http://arxiv.org/abs/2404.09446}
  {arXiv:2404.09446 [gr-qc]} \BibitemShut {NoStop}%
\bibitem [{\citenamefont {Lin}\ and\ \citenamefont {Chen}(2021)}]{Lin:2021bpe}%
  \BibitemOpen
  \bibfield  {author} {\bibinfo {author} {\bibfnamefont {Kuan-Nan}\
  \bibnamefont {Lin}}\ and\ \bibinfo {author} {\bibfnamefont {Pisin}\
  \bibnamefont {Chen}},\ }\bibfield  {title} {\enquote {\bibinfo {title}
  {{Particle production by a relativistic semitransparent mirror of finite size
  and thickness}},}\ }\href@noop {} {\  (\bibinfo {year} {2021})},\ \Eprint
  {http://arxiv.org/abs/2107.09033} {arXiv:2107.09033 [gr-qc]} \BibitemShut
  {NoStop}%
\bibitem [{\citenamefont {Kumar}\ \emph {et~al.}(2024)\citenamefont {Kumar},
  \citenamefont {Reyes},\ and\ \citenamefont {Wintergerst}}]{Kumar:2023kse}%
  \BibitemOpen
  \bibfield  {author} {\bibinfo {author} {\bibfnamefont {Piyush}\ \bibnamefont
  {Kumar}}, \bibinfo {author} {\bibfnamefont {Ignacio~A.}\ \bibnamefont
  {Reyes}}, \ and\ \bibinfo {author} {\bibfnamefont {Jakob}\ \bibnamefont
  {Wintergerst}},\ }\bibfield  {title} {\enquote {\bibinfo {title}
  {{Relativistic dynamics of moving mirrors in CFT2: Quantum backreaction and
  black holes}},}\ }\href {\doibase 10.1103/PhysRevD.109.065010} {\bibfield
  {journal} {\bibinfo  {journal} {Phys. Rev. D}\ }\textbf {\bibinfo {volume}
  {109}},\ \bibinfo {pages} {065010} (\bibinfo {year} {2024})},\ \Eprint
  {http://arxiv.org/abs/2310.03483} {arXiv:2310.03483 [hep-th]} \BibitemShut
  {NoStop}%
\bibitem [{\citenamefont {Reyes}(2021)}]{Reyes:2021npy}%
  \BibitemOpen
  \bibfield  {author} {\bibinfo {author} {\bibfnamefont {Ignacio~A.}\
  \bibnamefont {Reyes}},\ }\bibfield  {title} {\enquote {\bibinfo {title}
  {{Moving Mirrors, Page Curves, and Bulk Entropies in AdS2}},}\ }\href
  {\doibase 10.1103/PhysRevLett.127.051602} {\bibfield  {journal} {\bibinfo
  {journal} {Phys. Rev. Lett.}\ }\textbf {\bibinfo {volume} {127}},\ \bibinfo
  {pages} {051602} (\bibinfo {year} {2021})},\ \Eprint
  {http://arxiv.org/abs/2103.01230} {arXiv:2103.01230 [hep-th]} \BibitemShut
  {NoStop}%
\bibitem [{\citenamefont {Cozzella}\ \emph {et~al.}(2020)\citenamefont
  {Cozzella}, \citenamefont {Fulling}, \citenamefont {Landulfo},\ and\
  \citenamefont {Matsas}}]{Cozzella:2020gci}%
  \BibitemOpen
  \bibfield  {author} {\bibinfo {author} {\bibfnamefont {Gabriel}\ \bibnamefont
  {Cozzella}}, \bibinfo {author} {\bibfnamefont {Stephen~A.}\ \bibnamefont
  {Fulling}}, \bibinfo {author} {\bibfnamefont {Andr\'e G.~S.}\ \bibnamefont
  {Landulfo}}, \ and\ \bibinfo {author} {\bibfnamefont {George E.~A.}\
  \bibnamefont {Matsas}},\ }\bibfield  {title} {\enquote {\bibinfo {title}
  {{Uniformly accelerated classical sources as limits of Unruh-DeWitt
  detectors}},}\ }\href {\doibase 10.1103/PhysRevD.102.105016} {\bibfield
  {journal} {\bibinfo  {journal} {Phys. Rev. D}\ }\textbf {\bibinfo {volume}
  {102}},\ \bibinfo {pages} {105016} (\bibinfo {year} {2020})},\ \Eprint
  {http://arxiv.org/abs/2009.13246} {arXiv:2009.13246 [hep-th]} \BibitemShut
  {NoStop}%
\bibitem [{\citenamefont {Landulfo}\ \emph {et~al.}(2019)\citenamefont
  {Landulfo}, \citenamefont {Fulling},\ and\ \citenamefont
  {Matsas}}]{Landulfo:2019tqj}%
  \BibitemOpen
  \bibfield  {author} {\bibinfo {author} {\bibfnamefont {Andr\'e~G.S.}\
  \bibnamefont {Landulfo}}, \bibinfo {author} {\bibfnamefont {Stephen~A.}\
  \bibnamefont {Fulling}}, \ and\ \bibinfo {author} {\bibfnamefont
  {George~E.A.}\ \bibnamefont {Matsas}},\ }\bibfield  {title} {\enquote
  {\bibinfo {title} {{Classical and quantum aspects of the radiation emitted by
  a uniformly accelerated charge: Larmor-Unruh reconciliation and
  zero-frequency Rindler modes}},}\ }\href {\doibase
  10.1103/PhysRevD.100.045020} {\bibfield  {journal} {\bibinfo  {journal}
  {Phys. Rev. D}\ }\textbf {\bibinfo {volume} {100}},\ \bibinfo {pages}
  {045020} (\bibinfo {year} {2019})},\ \Eprint
  {http://arxiv.org/abs/1907.06665} {arXiv:1907.06665 [gr-qc]} \BibitemShut
  {NoStop}%
\bibitem [{\citenamefont {Fulling}\ \emph {et~al.}(2020)\citenamefont
  {Fulling}, \citenamefont {Landulfo},\ and\ \citenamefont
  {Matsas}}]{doi:10.1098/rspa.2020.0656}%
  \BibitemOpen
  \bibfield  {author} {\bibinfo {author} {\bibfnamefont {S.~A.}\ \bibnamefont
  {Fulling}}, \bibinfo {author} {\bibfnamefont {A.~G.~S.}\ \bibnamefont
  {Landulfo}}, \ and\ \bibinfo {author} {\bibfnamefont {G.~E.~A.}\ \bibnamefont
  {Matsas}},\ }\bibfield  {title} {\enquote {\bibinfo {title} {The relation
  between quantum and classical field theory with a classical source},}\ }\href
  {https://royalsocietypublishing.org/doi/abs/10.1098/rspa.2020.0656}
  {\bibfield  {journal} {\bibinfo  {journal} {Proceedings of the Royal Society
  A: Mathematical, Physical and Engineering Sciences}\ }\textbf {\bibinfo
  {volume} {476}},\ \bibinfo {pages} {20200656} (\bibinfo {year}
  {2020})}\BibitemShut {NoStop}%
\bibitem [{\citenamefont {Cozzella}\ \emph {et~al.}(2017)\citenamefont
  {Cozzella}, \citenamefont {Landulfo}, \citenamefont {Matsas},\ and\
  \citenamefont {Vanzella}}]{Cozzella:2017ckb}%
  \BibitemOpen
  \bibfield  {author} {\bibinfo {author} {\bibfnamefont {Gabriel}\ \bibnamefont
  {Cozzella}}, \bibinfo {author} {\bibfnamefont {Andr\'e G.~S.}\ \bibnamefont
  {Landulfo}}, \bibinfo {author} {\bibfnamefont {George E.~A.}\ \bibnamefont
  {Matsas}}, \ and\ \bibinfo {author} {\bibfnamefont {Daniel A.~T.}\
  \bibnamefont {Vanzella}},\ }\bibfield  {title} {\enquote {\bibinfo {title}
  {{Proposal for Observing the Unruh Effect using Classical
  Electrodynamics}},}\ }\href {\doibase 10.1103/PhysRevLett.118.161102}
  {\bibfield  {journal} {\bibinfo  {journal} {Phys. Rev. Lett.}\ }\textbf
  {\bibinfo {volume} {118}},\ \bibinfo {pages} {161102} (\bibinfo {year}
  {2017})},\ \Eprint {http://arxiv.org/abs/1701.03446} {arXiv:1701.03446
  [gr-qc]} \BibitemShut {NoStop}%
\bibitem [{\citenamefont {Ievlev}\ and\ \citenamefont
  {Good}(2023)}]{Ievlev:2023akh}%
  \BibitemOpen
  \bibfield  {author} {\bibinfo {author} {\bibfnamefont {Evgenii}\ \bibnamefont
  {Ievlev}}\ and\ \bibinfo {author} {\bibfnamefont {Michael R.~R.}\
  \bibnamefont {Good}},\ }\bibfield  {title} {\enquote {\bibinfo {title}
  {{Non-thermal photons and a Fermi-Dirac spectral distribution}},}\ }\href
  {\doibase 10.1016/j.physleta.2023.129131} {\bibfield  {journal} {\bibinfo
  {journal} {Phys. Lett. A}\ }\textbf {\bibinfo {volume} {488}},\ \bibinfo
  {pages} {129131} (\bibinfo {year} {2023})},\ \Eprint
  {http://arxiv.org/abs/2307.12860} {arXiv:2307.12860 [quant-ph]} \BibitemShut
  {NoStop}%
\bibitem [{\citenamefont {Stoney}(1881)}]{stoney1881physical}%
  \BibitemOpen
  \bibfield  {author} {\bibinfo {author} {\bibfnamefont {George~Johnstone}\
  \bibnamefont {Stoney}},\ }\bibfield  {title} {\enquote {\bibinfo {title} {On
  the physical units of nature},}\ }\href@noop {} {\bibfield  {journal}
  {\bibinfo  {journal} {Scientific Proceedings of the Royal Dublin Society}\
  }\textbf {\bibinfo {volume} {3}},\ \bibinfo {pages} {51--60} (\bibinfo {year}
  {1881})}\BibitemShut {NoStop}%
\bibitem [{\citenamefont {{Barrow}}(1983)}]{barrowSTONE}%
  \BibitemOpen
  \bibfield  {author} {\bibinfo {author} {\bibfnamefont {J.~D.}\ \bibnamefont
  {{Barrow}}},\ }\bibfield  {title} {\enquote {\bibinfo {title} {{Natural Units
  Before Planck}},}\ }\href
  {https://adsabs.harvard.edu/full/1983QJRAS..24...24B} {\bibfield  {journal}
  {\bibinfo  {journal} {Quarterly Journal of the Royal Astronomical Society}\
  }\textbf {\bibinfo {volume} {24}},\ \bibinfo {pages} {24--26} (\bibinfo
  {year} {1983})}\BibitemShut {NoStop}%
\bibitem [{\citenamefont {{International Bureau of Weights and
  Measures}}(2019)}]{SI}%
  \BibitemOpen
  \bibfield  {author} {\bibinfo {author} {\bibnamefont {{International Bureau
  of Weights and Measures}}},\ }\href
  {https://nvlpubs.nist.gov/nistpubs/SpecialPublications/NIST.SP.330-2019.pdf}
  {\emph {\bibinfo {title} {The International System of Units ({SI})}}},\
  \bibinfo {edition} {9th}\ ed.\ (\bibinfo {year} {2019})\BibitemShut {NoStop}%
\bibitem [{\citenamefont {Good}\ and\ \citenamefont
  {Linder}(2017)}]{Good:2017kjr}%
  \BibitemOpen
  \bibfield  {author} {\bibinfo {author} {\bibfnamefont {Michael R.~R.}\
  \bibnamefont {Good}}\ and\ \bibinfo {author} {\bibfnamefont {Eric~V.}\
  \bibnamefont {Linder}},\ }\bibfield  {title} {\enquote {\bibinfo {title}
  {{Slicing the Vacuum: New Accelerating Mirror Solutions of the Dynamical
  Casimir Effect}},}\ }\href {\doibase 10.1103/PhysRevD.96.125010} {\bibfield
  {journal} {\bibinfo  {journal} {Phys. Rev. D}\ }\textbf {\bibinfo {volume}
  {96}},\ \bibinfo {pages} {125010} (\bibinfo {year} {2017})},\ \Eprint
  {http://arxiv.org/abs/1707.03670} {arXiv:1707.03670 [gr-qc]} \BibitemShut
  {NoStop}%
\bibitem [{\citenamefont {Bianchi}\ and\ \citenamefont
  {Smerlak}(2014)}]{Bianchi:2014qua}%
  \BibitemOpen
  \bibfield  {author} {\bibinfo {author} {\bibfnamefont {Eugenio}\ \bibnamefont
  {Bianchi}}\ and\ \bibinfo {author} {\bibfnamefont {Matteo}\ \bibnamefont
  {Smerlak}},\ }\bibfield  {title} {\enquote {\bibinfo {title} {{Entanglement
  entropy and negative energy in two dimensions}},}\ }\href {\doibase
  10.1103/PhysRevD.90.041904} {\bibfield  {journal} {\bibinfo  {journal} {Phys.
  Rev. D}\ }\textbf {\bibinfo {volume} {90}},\ \bibinfo {pages} {041904}
  (\bibinfo {year} {2014})},\ \Eprint {http://arxiv.org/abs/1404.0602}
  {arXiv:1404.0602 [gr-qc]} \BibitemShut {NoStop}%
\bibitem [{\citenamefont {Barcelo}\ \emph {et~al.}(2011)\citenamefont
  {Barcelo}, \citenamefont {Liberati}, \citenamefont {Sonego},\ and\
  \citenamefont {Visser}}]{Barcelo:2010pj}%
  \BibitemOpen
  \bibfield  {author} {\bibinfo {author} {\bibfnamefont {Carlos}\ \bibnamefont
  {Barcelo}}, \bibinfo {author} {\bibfnamefont {Stefano}\ \bibnamefont
  {Liberati}}, \bibinfo {author} {\bibfnamefont {Sebastiano}\ \bibnamefont
  {Sonego}}, \ and\ \bibinfo {author} {\bibfnamefont {Matt}\ \bibnamefont
  {Visser}},\ }\bibfield  {title} {\enquote {\bibinfo {title} {{Minimal
  conditions for the existence of a Hawking-like flux}},}\ }\href {\doibase
  10.1103/PhysRevD.83.041501} {\bibfield  {journal} {\bibinfo  {journal} {Phys.
  Rev. D}\ }\textbf {\bibinfo {volume} {83}},\ \bibinfo {pages} {041501}
  (\bibinfo {year} {2011})},\ \Eprint {http://arxiv.org/abs/1011.5593}
  {arXiv:1011.5593 [gr-qc]} \BibitemShut {NoStop}%
\bibitem [{\citenamefont {Fern\'andez-Silvestre}\ \emph
  {et~al.}(2022)\citenamefont {Fern\'andez-Silvestre}, \citenamefont {Good},\
  and\ \citenamefont {Linder}}]{Fernandez-Silvestre:2022gqn}%
  \BibitemOpen
  \bibfield  {author} {\bibinfo {author} {\bibfnamefont {Diego}\ \bibnamefont
  {Fern\'andez-Silvestre}}, \bibinfo {author} {\bibfnamefont {Michael R.~R.}\
  \bibnamefont {Good}}, \ and\ \bibinfo {author} {\bibfnamefont {Eric~V.}\
  \bibnamefont {Linder}},\ }\bibfield  {title} {\enquote {\bibinfo {title}
  {{Upon the horizon\textquoteright{}s verge: Thermal particle creation between
  and approaching horizons}},}\ }\href {\doibase 10.1088/1361-6382/ac9d1b}
  {\bibfield  {journal} {\bibinfo  {journal} {Class. Quant. Grav.}\ }\textbf
  {\bibinfo {volume} {39}},\ \bibinfo {pages} {235008} (\bibinfo {year}
  {2022})},\ \Eprint {http://arxiv.org/abs/2208.01992} {arXiv:2208.01992
  [gr-qc]} \BibitemShut {NoStop}%
\bibitem [{\citenamefont {Good}\ and\ \citenamefont
  {Linder}(2018)}]{Good:2017ddq}%
  \BibitemOpen
  \bibfield  {author} {\bibinfo {author} {\bibfnamefont {Michael~R.R.}\
  \bibnamefont {Good}}\ and\ \bibinfo {author} {\bibfnamefont {Eric~V.}\
  \bibnamefont {Linder}},\ }\bibfield  {title} {\enquote {\bibinfo {title}
  {{Eternal and evanescent black holes and accelerating mirror analogs}},}\
  }\href {\doibase 10.1103/PhysRevD.97.065006} {\bibfield  {journal} {\bibinfo
  {journal} {Phys. Rev. D}\ }\textbf {\bibinfo {volume} {97}},\ \bibinfo
  {pages} {065006} (\bibinfo {year} {2018})},\ \Eprint
  {http://arxiv.org/abs/1711.09922} {arXiv:1711.09922 [gr-qc]} \BibitemShut
  {NoStop}%
\bibitem [{\citenamefont {Jackson}(1999)}]{Jackson:490457}%
  \BibitemOpen
  \bibfield  {author} {\bibinfo {author} {\bibfnamefont {John~David}\
  \bibnamefont {Jackson}},\ }\href {https://cds.cern.ch/record/490457} {\emph
  {\bibinfo {title} {{Classical electrodynamics; 3rd ed.}}}}\ (\bibinfo
  {publisher} {Wiley},\ \bibinfo {address} {New York, NY},\ \bibinfo {year}
  {1999})\BibitemShut {NoStop}%
\bibitem [{\citenamefont {Walker}\ and\ \citenamefont
  {Davies}(1982)}]{Walker_1982}%
  \BibitemOpen
  \bibfield  {author} {\bibinfo {author} {\bibfnamefont {W~R}\ \bibnamefont
  {Walker}}\ and\ \bibinfo {author} {\bibfnamefont {P~C~W}\ \bibnamefont
  {Davies}},\ }\bibfield  {title} {\enquote {\bibinfo {title} {An exactly
  soluble moving-mirror problem},}\ }\href {\doibase
  10.1088/0305-4470/15/9/008} {\bibfield  {journal} {\bibinfo  {journal}
  {Journal of Physics A: Mathematical and General}\ }\textbf {\bibinfo {volume}
  {15}},\ \bibinfo {pages} {L477--L480} (\bibinfo {year} {1982})}\BibitemShut
  {NoStop}%
\bibitem [{\citenamefont {Good}(2020)}]{good2020extreme}%
  \BibitemOpen
  \bibfield  {author} {\bibinfo {author} {\bibfnamefont {Michael~R.R.}\
  \bibnamefont {Good}},\ }\bibfield  {title} {\enquote {\bibinfo {title}
  {{Extremal Hawking radiation}},}\ }\href {\doibase
  10.1103/PhysRevD.101.104050} {\bibfield  {journal} {\bibinfo  {journal}
  {Phys. Rev. D}\ }\textbf {\bibinfo {volume} {101}},\ \bibinfo {pages}
  {104050} (\bibinfo {year} {2020})},\ \Eprint
  {http://arxiv.org/abs/2003.07016} {arXiv:2003.07016 [gr-qc]} \BibitemShut
  {NoStop}%
\bibitem [{\citenamefont {Good}\ \emph {et~al.}(2020)\citenamefont {Good},
  \citenamefont {Linder},\ and\ \citenamefont {Wilczek}}]{Good:2019tnf}%
  \BibitemOpen
  \bibfield  {author} {\bibinfo {author} {\bibfnamefont {Michael~R.R.}\
  \bibnamefont {Good}}, \bibinfo {author} {\bibfnamefont {Eric~V.}\
  \bibnamefont {Linder}}, \ and\ \bibinfo {author} {\bibfnamefont {Frank}\
  \bibnamefont {Wilczek}},\ }\bibfield  {title} {\enquote {\bibinfo {title}
  {{Moving mirror model for quasithermal radiation fields}},}\ }\href {\doibase
  10.1103/PhysRevD.101.025012} {\bibfield  {journal} {\bibinfo  {journal}
  {Phys. Rev. D}\ }\textbf {\bibinfo {volume} {101}},\ \bibinfo {pages}
  {025012} (\bibinfo {year} {2020})},\ \Eprint
  {http://arxiv.org/abs/1909.01129} {arXiv:1909.01129 [gr-qc]} \BibitemShut
  {NoStop}%
\bibitem [{\citenamefont {Carlitz}\ and\ \citenamefont
  {Willey}(1987{\natexlab{b}})}]{carlitz1987reflections}%
  \BibitemOpen
  \bibfield  {author} {\bibinfo {author} {\bibfnamefont {Robert~D.}\
  \bibnamefont {Carlitz}}\ and\ \bibinfo {author} {\bibfnamefont {Raymond~S.}\
  \bibnamefont {Willey}},\ }\bibfield  {title} {\enquote {\bibinfo {title}
  {Reflections on moving mirrors},}\ }\href {\doibase 10.1103/PhysRevD.36.2327}
  {\bibfield  {journal} {\bibinfo  {journal} {Phys. Rev. D}\ }\textbf {\bibinfo
  {volume} {36}},\ \bibinfo {pages} {2327--2335} (\bibinfo {year}
  {1987}{\natexlab{b}})}\BibitemShut {NoStop}%
\end{thebibliography}%
\end{document}